\documentclass[preprint,times,5p,twocolumn]{elsarticle}
\usepackage[ansinew]{inputenc}

\usepackage{amsmath}
\usepackage{epic}
\usepackage{wrapfig}
\usepackage{eepic}
\usepackage{latexsym}
\usepackage{array}
\usepackage{multicol}

\usepackage{fancyhdr}
\usepackage{verbatim}
\usepackage[colorlinks=true, pdfstartview=FitV, linkcolor=blue, citecolor=blue, urlcolor=blue, unicode]{hyperref}
\usepackage{ifpdf}
\usepackage{cite}

\usepackage{enumitem}

\usepackage{graphicx}
\graphicspath{{figures/}}

\journal{Computer Physics Communications}

\begin{document}

\begin{frontmatter}

\title{Delphes, a framework for fast simulation of a generic collider experiment}
\author{S. Ovyn\corref{cor1}}
\ead{severine.ovyn@uclouvain.be}

\author{X. Rouby}
\author{V. Lema\^itre}

\address{Center for Particle Physics and Phenomenology (CP3),\\
        Universit\'e catholique de Louvain,\\
        B-1348 Louvain-la-Neuve, Belgium}

\begin{abstract}

This paper presents a new \texttt{C++} framework, \textit{Delphes}, performing a
fast multipurpose detector response simulation.
The simulation includes a tracking system, embedded into a magnetic field,
calorimeters and a muon system, and possible very forward detectors arranged
along the beamline.
The framework is interfaced to standard file formats (e.g.\ Les Houches Event
File or \texttt{HepMC}) and outputs observables such as isolated leptons,
missing transverse energy and collection of jets which can be used for dedicated
analyses. The simulation of the detector response takes into account the effect
of magnetic field, the granularity of the calorimeters and subdetector
resolutions.
A simplified preselection can also be applied on processed events for trigger
emulation. Detection of very forward scattered particles relies on the transport
in beamlines with the \textit{Hector} software. Finally, the \textsc{FROG} 2D/3D
event display is used for visualisation of the collision final states.
\\ \\

\textit{Preprint:} \texttt{CP3-09-01}, \texttt{arXiv:0903.2225 [hep-ph]}\\ \\
\includegraphics[scale=0.8]{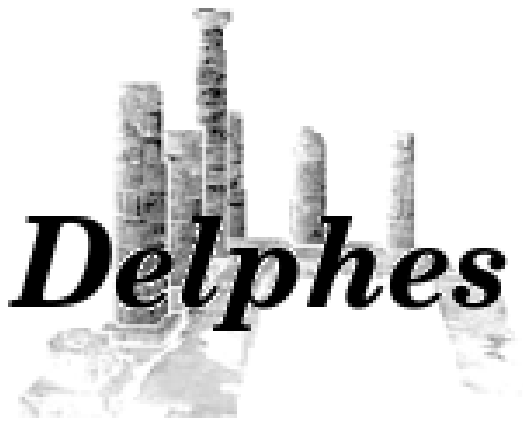}\\
{\bf PROGRAM SUMMARY}\\
\begin{small}
\noindent
{\em Program Title:} DELPHES                                  \\
{\em Current version:} 1.8                                    \\
{\em Journal Reference:}                                      \\
{\em Catalogue identifier:}                                   \\
{\em Distribution format:}  tar.gz                            \\
{\em Programming language:}  C++                              \\
{\em External routines/libraries:} ROOT environment \\
{\em Subprograms used:} HepMC, StdHEP, FASTJET, \textit{Hector}, FROG. All provided within \textit{Delphes} distribution.  \\
{\em URL:}\href{http://www.fynu.ucl.ac.be/delphes.html}{http://www.fynu.ucl.ac.be/delphes.html}\\
\end{small}


\end{abstract}
\cortext[cor1]{Corresponding author: +32.10.47.32.29.}
\end{frontmatter}

\section{Introduction}

Multipurpose detectors at high energy colliders are very complex systems. 
Precise data analyses require a full detector simulation, including transport of
the primary and secondary particles through the detector material accounting for
the various detector inefficiencies, the dead material, the imperfections and
the geometrical details. Their simulation is in general performed by means of
the GEANT~\citep{bib:geant} package and final observables used for analyses
usually require sophisticated reconstruction algorithms.

This complexity can only be handled by large collaborations.  
Phenomenological studies, looking for the observability of given signals, 
require in general only fast but realistic estimates of the expected signal
signatures and their associated backgrounds.

In this context, a new framework, called \textit{Delphes}~\citep{bib:delphes},
has been developped, for a fast simulation of a general-purpose collider
experiment.
Using this framework, observables such as cross-sections and efficiencies after
event selection can be estimated for specific reactions.
Starting from the output of event generators, the simulation of the detector
response takes into account the subdetector resolutions, by smearing the
kinematics of final-state particles (i.e. those considered as stable by the
event generator
\footnote{In the current \textit{Delphes} version, particles other than
electrons ($e^\pm$), photons ($\gamma$), muons ($\mu^\pm$), neutrinos ($\nu_e$,
$\nu_\mu$ and $\nu_\tau$) and neutralinos are simulated as hadrons for their
interactions with the calorimeters. The simulation of stable particles beyond
the Standard Model should therefore be handled with
care~\citep{qr:invisibleparticles}.}).

\begin{figure*}[!ht]
\begin{center}
\includegraphics[scale=0.78]{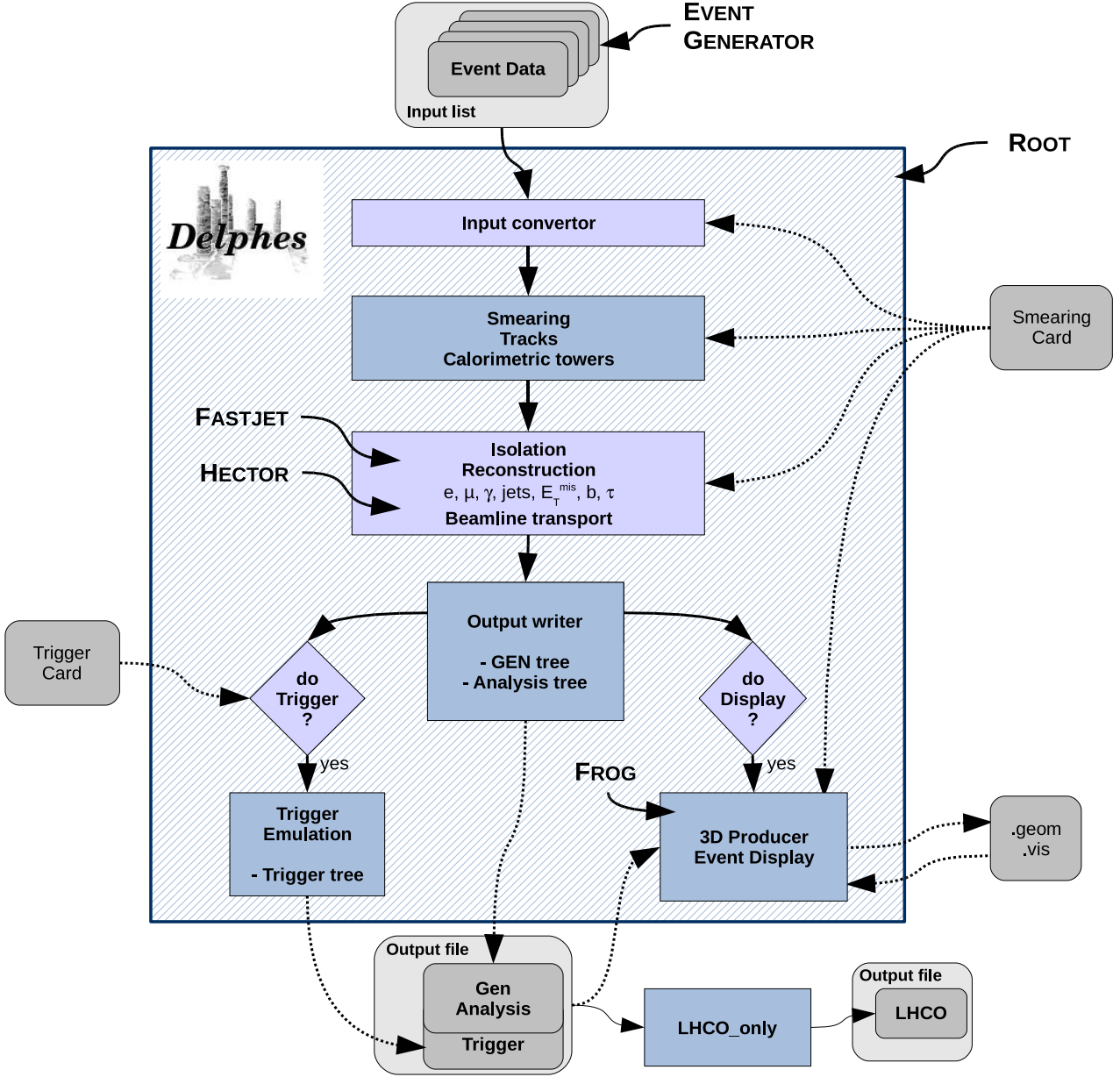}
\caption{Flow chart describing the principles behind \textit{Delphes}. Event
files coming from external Monte Carlo generators are read by a converter stage
(top).
The kinematics variables of the final-state particles are then smeared
according to the tunable subdetector resolutions. 
Tracks are reconstructed in a simulated solenoidal magnetic field and
calorimetric cells sample the energy deposits. Based on these low-level objects,
dedicated algorithms are applied for particle identification, isolation and
reconstruction. 
The transport of very forward particles to the near-beam detectors is also
simulated. 
Finally, an output file is written, including generator-level and
analysis-object data. 
If requested, a fully parametrisable trigger can be emulated. Optionally, the
geometry and visualisation files for the 3D event display can also be produced.
All user parameters are set in the \textit{Detector/Smearing Card} and the
\textit{Trigger Card}. }
\label{fig:FlowChart}
\end{center}
\end{figure*}

\textit{Delphes} includes the most crucial experimental features, such as
(Fig.~\ref{fig:FlowChart}):
\begin{enumerate}
\item the geometry of both central and forward detectors,
\item the effect of magnetic field on tracks,
\item the reconstruction of photons, leptons, jets, $b$-jets, $\tau$-jets and
missing transverse energy,
\item a lepton isolation,
\item a trigger emulation,
\item an event display.
\end{enumerate}

Although \textit{Delphes} yields much realistic results than a simple
``parton-level" analysis, it has some limitations. Detector geometry is
idealised, being uniform, symmetric around the beam axis, and having no cracks
nor dead material. Secondary interactions, multiple scatterings, photon
conversion and bremsstrahlung are also neglected.

Several common datafile formats can be used as input in \textit{Delphes}
\citep{qr:inputformat}, in order to process events from many different
generators. \textit{Delphes} creates output data in a ROOT ntuple
\citep{bib:Root}. This output contains a copy of the generator-level data, the
analysis data objects after reconstruction, and possibly the results of the
trigger emulation \citep{qr:outputformat}.
In option \textit{Delphes} can produce a reduced output file in \texttt{*.lhco}
text format, which is limited to the list of the reconstructed high-level
objects in the final states~\citep{qr:lhco}.

\section{Simulation of the detector response}

\begin{figure}[!ht]
\begin{center}
\includegraphics[width=\columnwidth]{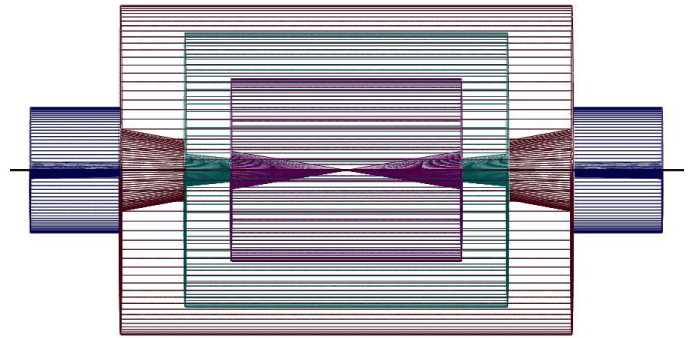}
\caption{
Profile of layout of the generic detector geometry assumed in \textit{Delphes}.
The innermost layer, close to the interaction point, is a central tracking
system (pink). It is surrounded by a central calorimeter volume (green) with
both electromagnetic and hadronic sections. The outer layer of the central
system (red) is muon system. In addition, two end-cap calorimeters (blue) extend
the pseudorapidity coverage of the central detector.
Additional forward detectors are not depicted.
}
\label{fig:GenDet3}
\end{center}
\end{figure}

The overall layout of the multipurpose detector simulated by \textit{Delphes}
is shown in Fig.~\ref{fig:GenDet3}. It consists in a central tracking system
(\textsc{TRACKER}) surrounded by an electromagnetic and a hadron calorimeters
(\textsc{ECAL} and \textsc{HCAL}, each with a central region and two endcaps)
and two forward calorimeters (\textsc{FCAL}). Finally, a muon system
(\textsc{MUON}) encloses the central detector volume.

A detector card \citep{qr:detectorcard} allows a large spectrum of running
conditions by modifying basic detector parameters, including calorimeter and
tracking coverage and resolution, thresholds or jet algorithm parameters. 
Even if \textit{Delphes} has been developped for the simulation of
general-purpose detectors at the \textsc{LHC} (namely, \textsc{CMS} and
\textsc{ATLAS}), this input parameter file interfaces a flexible parametrisation
for other cases, e.g.\ at future linear colliders~\citep{qr:datacards}.
The geometrical coverage of the various subsystems used in the default
configuration are summarised in Tab.~\ref{tab:defEta}. 

\begin{table}[t]
\begin{center}
\caption{Default extension in pseudorapidity $\eta$ of the different subdetectors. 
Full azimuthal ($\phi$) acceptance is assumed.
 \vspace{0.5cm}}
\begin{tabular}{lcc}
\hline
 & $\eta$ & $\phi$ \\ \hline
\textsc{TRACKER} 	& $[-2.5; 2.5]$ 		& $[-\pi ; \pi]$\\
\textsc{ECAL}, \textsc{HCAL} & $[-1.7 ; 1.7]$		& $[-\pi ; \pi]$\\
\textsc{ECAL}, \textsc{HCAL} endcaps & $[-3 ; -1.7]$ \& $[1.7 ; 3]$	& $[-\pi ; \pi]$\\
\textsc{FCAL} 		& $[-5 ; -3]$ \& $[3 ;5]$     	& $[-\pi ; \pi]$\\
\textsc{MUON} 		& $[-2.4 ; 2.4]$ 		& $[-\pi ; \pi]$\\ \hline
\end{tabular}
\label{tab:defEta}
\end{center}
\end{table}

\subsection{Magnetic field}
In addition to the subdetectors, the effects of a solenoidal magnetic field are
simulated for the charged particles~\citep{qr:magneticfield}. This affects the
position at which charged particles enter the calorimeters and their
corresponding tracks. The field extension is limited to the tracker volume and
is in particular not applied for muon chambers. This is not a limiting
factor since the magnetic field is not used for the muon momentum smearing.

\subsection{Tracks reconstruction}
Every stable charged particle with a transverse momentum above some threshold
and lying inside the detector volume covered by the tracker provides a track. 
By default, a track is assumed to be reconstructed with $90\%$ probability if
its transverse momentum $p_T$ is higher than $0.9~\textrm{GeV}/c$ and if its
pseudorapidity $|\eta| \leq 2.5$~\citep{qr:tracks}. No smearing is currently
applied on track parameters. For each track, the positions at vertex
$(\eta,\phi)$ and at the entry point in the calorimeter layers
$(\eta,\phi)_{calo}$ are available.

\subsection{Calorimetric cells}

The response of the calorimeters to energy deposits of incoming particles
depends on their segmentation and resolution, as well as on the nature of the
particles themselves. In CMS and ATLAS detectors, for instance, the calorimeter
characteristics are not identical in every direction, with typically finer
resolution and granularity in the central
regions~\citep{bib:cmsjetresolution,bib:ATLASresolution}. It is thus very
important to compute the exact coordinates of the entry point of the particles
into the calorimeters, in taking the magnetic field effect into account.

\begin{figure}[!ht]
\begin{center}
\includegraphics[width=\columnwidth]{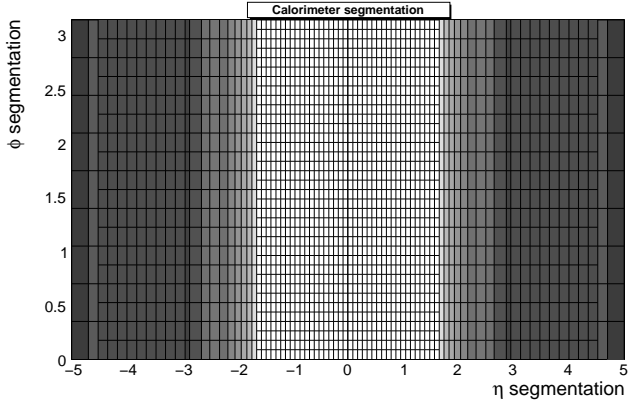}
\caption{Default segmentation of the calorimeters in the $(\eta,\phi)$ plane. Only the central detectors (\textsc{ECAL}, \textsc{HCAL}) and \textsc{FCAL} are considered. $\phi$ angles are expressed in radians.}
\label{fig:calosegmentation}
\end{center}
\end{figure}

The smallest unit for geometrical sampling of the calorimeters is a
\textit{cell}; it segments the $(\eta,\phi)$ plane for the energy measurement.
No longitudinal segmentation is available in the simulated
calorimeters. \textit{Delphes} assumes that ECAL and HCAL have the same
segmentations and that the detector is symmetric in $\phi$ and with respect to
the $\eta=0$ plane~\citep{qr:calorimetriccells}.
Fig.~\ref{fig:calosegmentation} illustrates the default calorimeter
segmentation.

The calorimeter response is parametrised through a Gaussian smearing of the accumulated cell energy with a variance $\sigma$:
\begin{equation} 
\frac{\sigma}{E} = \frac{S}{\sqrt{E}} \oplus \frac{N}{E} \oplus C,
\label{eq:caloresolution}
\end{equation}
where $S$, $N$ and $C$ are the \textit{stochastic}, \textit{noise} and \textit{constant} terms, respectively, and $\oplus$ stands for quadratic additions~\citep{qr:energysmearing}.\\

In the default parametrisation, ECAL and HCAL are assumed to cover the
pseudorapidity range $|\eta|<3$, and FCAL between $3.0$ and $5.0$, with
different response to electrons and photons, or to hadrons. 
Muons and neutrinos are assumed not to interact with the
calorimeters~\citep{qr:invisibleparticles}. The default values of the
stochastic, noise and constant terms are given in Tab.~\ref{tab:defResol}.\\

\begin{table}[!h]
\begin{center}
\caption{Default values for the resolution of the central and forward
calorimeters (for both electromagnetic and hadronic parts). Resolution is
parametrised by the \textit{stochastic} ($S$), \textit{noise} ($N$) and
\textit{constant} ($C$) terms
(Eq.~\ref{eq:caloresolution})~\citep{qr:resolutionterms}.
\vspace{0.5cm}}
\begin{tabular}[!h]{lccc}
\hline
        & $S$ (GeV$^{1/2}$) & $N$ (GeV) & $C$ \\\hline
  ECAL            & $0.05$   & $0.25$ & $0.0055$ \\
  ECAL, end caps  & $0.05$  & $0.25$ & $0.0055$ \\
  FCAL, e.m. part & $2.084$ & $0$    & $0.107$ \\
  HCAL            & $1.5$   & $0$    & $0.05$\\
  HCAL, end caps  & $1.5$   & $0$    & $0.05$\\
  FCAL, had. part & $2.7$   & $0$    & $0.13$\\
\hline
\end{tabular}
\label{tab:defResol}
\end{center}
\end{table}

Electrons and photons are assumed to leave their energy in the electromagnetic
parts of the calorimeters (\textsc{ECAL} and \textsc{FCAL}, e.m.), while charged
and neutral final-state hadrons are assumed to leave their entire energy
interactin the hadronic parts (\textsc{HCAL} and \textsc{FCAL}, had.).
Some long-living particles, such as the $K^0_s$ and $\Lambda$'s, with lifetime
$c\tau$ smaller than $10~\textrm{mm}$ are considered as stable particles by the
generators although they may decay before reaching the calorimeters. The energy
smearing of such particles is therefore performed using the expected fraction of
the energy, determined according to their decay products, that would be
deposited into the \textsc{ECAL} ($E_{\textsc{ECAL}}$) and into the
\textsc{HCAL} ($E_{\textsc{HCAL}}$). Defining $F$ as the fraction of the energy
leading to a \textsc{HCAL} deposit, the two energy values are given by
\begin{equation}
\left\{
\begin{array}{l}
E_{\textsc{HCAL}} = E \times F \\
E_{\textsc{ECAL}} = E \times (1-F) \\
\end{array}
\right.
\end{equation}
where $0 \leq F \leq 1$. The resulting calorimetry energy measurement given
after the application of the smearing is then $E = E_{\textsc{HCAL}} +
E_{\textsc{ECAL}}$. For $K_S^0$ and $\Lambda$ hadrons, the energy fraction is
$F$ is assumed to be $0.7$~\citep{qr:emhadratios}.\\

No sharing between neighbouring cells is implemented when particles enter a
cell very close to its geometrical edge. Due to the finite segmentation, the
smearing, as defined in Eq.~\ref{eq:caloresolution}, is applied directly on the
accumulated electromagnetic and hadronic energies of each calorimetric cell. The
calorimetric cells enter in the calculation of the missing transverse energy
(\textsc{MET}), and are used as input for the jet reconstruction algorithms.

The output file created by \textit{Delphes}~\citep{qr:analysistree} stores the
final collections of particles ($e^\pm$, $\mu^\pm$, $\gamma$) and objects (light
jets, $b$-jets, $\tau$-jets, $E_T^\textrm{miss}$). In addition, collections of
tracks, calorimetric cells and hits in the very forward detectors (\textsc{ZDC},
\textsc{RP220} and \textsc{FP420}, see Sec.~\ref{sec:vfd}) are added. 

\section{High-level reconstruction}

While electrons,
muons and photons are easily identified, other quantities are more difficult to
evaluate as they rely on sophisticated algorithms (e.g. jets or missing energy).

For most of these objects, their four-momentum and related quantities are
directly accessible in \textit{Delphes} output ($E$, $\vec{p}$, $p_T$, $\eta$
and $\phi$). Additional properties are available for specific objects (like the
charge and the isolation status for $e^\pm$ and $\mu^\pm$, the result of
application of $b$-tag for jets and time-of-flight for some detector hits).

\subsection{Photon and charged lepton}
From here onwards, \textit{electrons} refer to both positrons ($e^+$) and
electrons ($e^-$), and $\textit{charged leptons}$ refer to electrons and muons
($\mu^\pm$), leaving out the $\tau^\pm$ leptons as they decay before being
detected. 

The electron, muon and photon collections contains only the true final-state
particles identified via the generator-data. In addition, these particles must
pass fiducial cuts taking into account the magnetic field effects and some
additional reconstruction cuts.

Consequently, no fake candidates enter these collections. However, when needed,
fake candidates can be added into the collections at the analysis level, when
processing \textit{Delphes} output data. As effects like bremsstrahlung are not
taken into account along the lepton propagation in the tracker, no clustering is
needed for the electron reconstruction in \textit{Delphes}.

\subsubsection*{Electrons and photons}
Real electron ($e^\pm$) and photon candidates are associated to the final-state
collections if they fall into the acceptance of the tracking system and have a
transverse momentum above some threshold (default: $p_T > 10~\textrm{GeV}/c$). 
\textit{Delphes} assumes a perfect
algorithm for clustering and Brehmstrahlung recovery. Electron energy is smeared
according to the resolution of the calorimetric cell where it points to, but
independently from any other deposited energy in this cell. 
Electrons and photons may create a candidate in the jet collection. The $(\eta,
\phi)$ position at vertex corresponds to corresponding track vertex.

\subsubsection*{Muons}
Generator-level muons entering the muon detector acceptance (default: $-2.4
\leq \eta \leq 2.4$) and overpassing some threshold (default: $p_T >
10~\textrm{GeV}/c$) are considered as good candidates for analyses.
The application of the detector resolution on the muon momentum depends on a
Gaussian smearing of the $p_T$~\citep{qr:muonsmearing}.
Neither $\eta$ nor $\phi$ variables are modified beyond the calorimeters.
Multiple scattering is neglected. This implies that low energy muons have in
\textit{Delphes} a better resolution than in a real detector.  At last, the
particles which might leak out of the calorimeters into the muon systems
(\textit{punch-through}) are not considered as muon candidates in
\textit{Delphes}.

\subsubsection*{Charged lepton isolation}
\label{sec:isolation}

To improve the quality of the contents of the charged lepton collections,
isolation criteria can be applied. This requires that electron or muon
candidates are isolated in the detector from any other particle, within a small
cone. In \textit{Delphes}, charged lepton isolation demands by default that
there is no other charged particle with $p_T>2~\textrm{GeV}/c$ within a cone of
$\Delta R = \sqrt{\Delta \eta^2 + \Delta \phi^2} <0.5$ centered on the cell
associated to the charged lepton $\ell$, obviously taking the magnetic field
into account.

The result (i.e.\ \textit{isolated} or \textit{not}) is added to the charged lepton measured properties.
In addition, the sum $P_T$ of the transverse momenta of all tracks but the lepton one within the isolation cone is 
provided~\citep{qr:isolflag}:
$$ P_T = \sum_{i \neq \ell}^\textrm{tracks} p_T(i)$$

No calorimetric isolation is applied, but the charged lepton collections
contain also the ratio $\rho_\ell$ between (1) the sum of the transverse
energies in all calorimetric cells in a $N \times N$ grid around the lepton, and
(2) the lepton transverse momentum~\citep{qr:caloisolation}:
$$ \rho_\ell = \frac{\Sigma_i E_T(i)}{p_T(\ell)}~,~ i\textrm{ in }N \times N
\textrm { grid centred on }\ell.$$

\subsection{Jet reconstruction}

A realistic analysis requires a correct treatment of partons which have
hadronised. Therefore, the most widely currently used jet algorithms have been
integrated into the \textit{Delphes} framework using the FastJet
tools\footnote{A more detailed description of the jet algorithms is given in the
User Manual, in appendix.}. Six different jet reconstruction schemes are
available~\citep{bib:FASTJET,qr:jetalgo}. For all of them, the calorimetric
cells are used as inputs. Jet algorithms differ in their sensitivity to soft
particles or collinear splittings, and in their computing speed performances.
 
\subsubsection*{Cone algorithms}
 
\begin{enumerate}
 
\item {\it CDF Jet Clusters}~\citep{bib:jetclu}: Cone algorithm forming jets by
combining cells lying within a circle (default radius $\Delta R=0.7$) in the
$(\eta$, $\phi)$ space. Jets are seeded by all cells with transverse energy
$E_T$ above a given threshold (default: $E_T >
1~\textrm{GeV}$)~\citep{qr:jetparams}. 
 
\item {\it CDF MidPoint}~\citep{bib:midpoint}: Cone algorithm with additional
``midpoints'' (energy barycentres) in the list of seeds.
 
\item {\it Seedless Infrared Safe Cone}~\citep{bib:SIScone}: The
\textsc{SISC}one algorithm is simultaneously insensitive to additional soft
particles and collinear splittings.
\end{enumerate}

\subsubsection*{Recombination algorithms}

The next three jet algorithms rely on recombination schemes where calorimeter
cell pairs are successively merged:
 
\begin{enumerate}[start=4]
\item {\it Longitudinally invariant $k_t$ jet}~\citep{bib:ktjet},
\item {\it Cambridge/Aachen jet}~\citep{bib:aachen},
\item {\it Anti $k_t$ jet}~\citep{bib:antikt}, where hard jets are exactly
circular in the $(y,\phi)$ plane.
\end{enumerate}

The recombination algorithms are safe with respect to soft radiations
(\textit{infrared}) and collinear splittings. Their implementations are similar
except for the definition of the \textit{distances} used during the merging
procedure. 

By default, reconstruction uses the CDF cone algorithm. Jets are stored if their
transverse energy is higher than $20~\textrm{GeV}$~\citep{qr:ptcutjet}.

\subsubsection*{Energy flow}

In jets, several particle can leave their energy into a given calorimetric cell,
which broadens the jet energy resolution. However, the energy of charged
particles associated to jets can be deduced from their associated track, thus
providing a way to identify some of the components of cells with multiple hits.
When the \textit{energy flow} is switched on in \textit{Delphes}, the energy of
tracks pointing to calorimetric cells is subtracted and smeared separately,
before running the chosen jet reconstruction algorithm. This option allows a
better jet energy reconstruction~\citep{qr:energyflow}.
 
\subsection{$b$-tagging}
\label{btagging}

A jet is tagged as $b$-jets if its direction lies in the acceptance of the
tracker and if it is associated to a parent $b$-quark. 
The (mis)tagging relies on the identity of
the most energetic parton within a cone around the jet axis, with a
radius equal to the one used to reconstruct the jet (default: $\Delta R$ of
$0.7$). 
By default, a $b$-tagging efficiency of $40\%$ is assumed if the jet has a
parent $b$ quark. For $c$-jets and light jets (i.e.\ originating in $u$, $d$,
$s$ quarks or in gluons), a fake $b$-tagging efficiency of $10 \%$ and $1 \%$
is assumed respectively~\citep{qr:btag}. Therefore, in current version of
\textit{Delphes}, the displacement of secondary vertices is not taken into
account. As such, the $b$-tagging efficiency is below the expected $40\%$.

\subsection{Identification of hadronic \texorpdfstring{$\tau$}{\texttau} decays}

Jets originating from $\tau$-decays are identified using a procedure consistent
with the one applied in a full detector simulation~\citep{bib:cmsjetresolution}.
The tagging relies on two properties of the $\tau$ lepton. First, $77\%$ of the
$\tau$ hadronic decays contain only one charged hadron associated to a few
neutrals (\textit{1-prong}). Secondly, the particles arisen from the
$\tau$ lepton produce narrow jets in the calorimeter (this is defined as the jet
\textit{collimation}). 

\begin{figure}[!ht]
\begin{center}
\includegraphics[width=0.80\columnwidth]{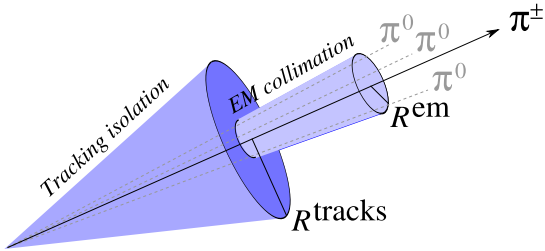}
\caption{Illustration of the identification of $\tau$-jets ($1-$prong). The jet cone is narrow and contains only one track. The small cone serves to apply the \textit{electromagnetic collimation}, while the broader cone is used to reconstruct the jet originating from the $\tau$-decay.}
\label{h_WW_ss_cut1}
\end{center}
\end{figure}

\begin{table}[!h]
\begin{center}
\caption{Default values for parameters used in $\tau$-jet reconstruction algorithm. Electromagnetic collimation requirements involve the inner \textit{small} cone radius $R^\textrm{em}$, the minimum transverse energy for calorimetric cells $E_T^\textrm{cell}$ and the collimation factor $C_\tau$. Tracking isolation constrains the number of tracks with a significant transverse momentum $p_T^\textrm{tracks}$ in a cone of radius $R^\textrm{tracks}$.  Finally, the $\tau$-jet collection is purified by the application of a cut on the $p_T$ of $\tau$-jet candidates~\citep{qr:taujets}.
\vspace{0.5cm}  }
\begin{tabular}[!h]{lll}
\hline
\multicolumn{3}{l}{\textbf{Electromagnetic collimation}} \\
& $R^\textrm{em}$      	   & $0.15$\\
& min $E_{T}^\textrm{cell}$ & $1.0$~GeV\\
& $C_{\tau}$           	   & $0.95$\\
\multicolumn{3}{l}{\textbf{Tracking isolation}} \\
& $R^\textrm{tracks}$  	   & $0.4$\\
& min $p_T^\textrm{tracks}$  & $2$ GeV$/c$\\
\multicolumn{3}{l}{\textbf{$\tau$-jet candidate}} \\
& $\min p_T$ 		   & $10$ GeV$/c$\\
\hline
\end{tabular}
\label{tab:tauRef}
\end{center}
\end{table}

\subsubsection*{Electromagnetic collimation}

\begin{figure}[!ht]
\begin{center}
\includegraphics[width=\columnwidth]{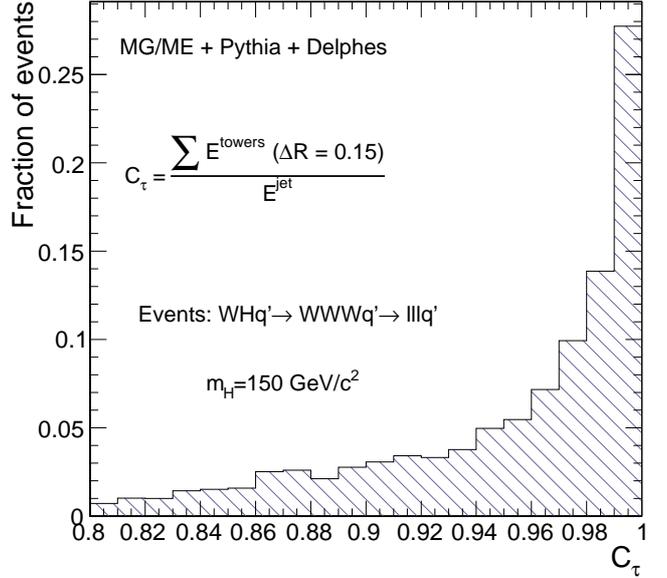}
\caption{Distribution of the electromagnetic collimation $C_\tau$ variable for true $\tau$-jets, normalised to unity. This distribution is shown for associated $WH$ photoproduction~\citep{bib:whphotoproduction}, where the Higgs boson decays into a $W^+ W^-$ pair. Each $W$ boson decays into a $\ell \nu_\ell$ pair, where $\ell = e, \mu, \tau$.
Events generated with MadGraph/MadEvent~\citep{bib:mgme}.
Final state hadronisation is performed by \textit{Pythia}~\citep{bib:pythia}.
Histogram entries correspond to true $\tau$-jets, matched with generator-level data. }
\label{fig:tau2}
\end{center}
\end{figure}

To use the narrowness of the $\tau$-jet, the \textit{electromagnetic
collimation} $C_{\tau}$ is defined as the sum of the energy of cells in a small
cone of radius $R^\textrm{em}$ around the jet axis, divided by the energy of the
reconstructed jet. To be taken into account, a calorimeter cell should have a
transverse energy $E_T^\textrm{cell}$ above a given threshold. A large fraction
of the jet energy is expected in this small cone. This fraction, or
\textit{collimation factor}, is represented in Fig.~\ref{fig:tau2} for the
default values (see Tab.~\ref{tab:tauRef}).

\subsubsection*{Tracking isolation}

The tracking isolation for the $\tau$ identification requires that the number
of tracks associated to particles with significant transverse momenta is one and
only one in a cone of radius $R^\textrm{tracks}$ ($3-$prong $\tau$-jets are
rejected). This cone should be entirely incorporated into the tracker to be
taken into account. Default values of these parameters are given in
Tab.~\ref{tab:tauRef}.

\begin{figure}[!ht]
\begin{center}
\includegraphics[width=\columnwidth]{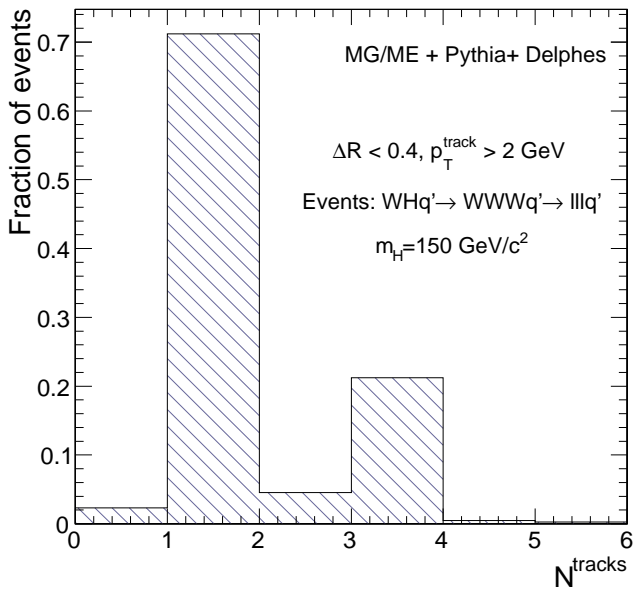}
\caption{Distribution of the number of tracks $N^\textrm{tracks}$ within a small jet cone for true $\tau$-jets, normalised to unity. Photoproduced $WH$ events, where $W$ bosons decay leptonically ($e,\mu,\tau$), as in Fig.~\ref{fig:tau2}. 
Histogram entries correspond to true $\tau$-jets, matched with generator-level data.}
\label{fig:tau1}
\end{center}
\end{figure}

\subsubsection*{Purity}
Once both electromagnetic collimation and tracking isolation are applied, a
threshold on the $p_T$ of the $\tau$-jet candidate is requested to purify the
collection. This procedure selects $\tau$ leptons decaying hadronically with a
typical efficiency of $66\%$. 

\subsection{Missing transverse energy}
In an ideal detector, momentum conservation imposes the transverse momentum of
the observed final state $\overrightarrow{p_T}^\textrm{obs}$ to be equal and
in opposite direction to the $\overrightarrow{p_T}$ vector sum of the
invisible particles, written $\overrightarrow{p_T}^\textrm{miss}$.
The \textit{true} missing transverse energy, i.e.\ at generator-level, is
calculated as the opposite of the vector sum of the transverse momenta of all
visible particles -- or equivalently, to the vector sum of invisible particle
transverse momenta. 
In a real experiment, calorimeters measure energy and not momentum. Any problem
affecting the detector (dead channels, misalignment, noisy cells, cracks)
worsens directly the measured missing transverse energy $\overrightarrow
{E_T}^\textrm{miss}$. In \textit{Delphes}, \textsc{MET} is based on the
calorimetric cells only. Muons and neutrinos are therefore
not taken into account for its evaluation:
\begin{equation}
\overrightarrow{E_T}^\textrm{miss} = - \sum^\textrm{cells}_i \overrightarrow{E_T}(i)
\end{equation}
However, as muon candidates, tracks and calorimetric cells are available in the
output file, the missing transverse energy can always be reprocessed a
posteriori with more specialised algorithms.

\section{Trigger emulation}

Most of the usual trigger algorithms select events containing leptons, jets, and
\textsc{MET} with an energy scale above some threshold. This is often expressed
in terms of a cut on the transverse momentum of one or several objects of the
measured event. Logical combinations of several conditions are also possible.
For instance, a trigger path could select events containing at least one jet and
one electron such as $p_T^\textrm{jet} > 100~\textrm{GeV}/c$ and $p_T^e >
50~\textrm{GeV}/c$.

A trigger emulation is included in \textit{Delphes}, using a fully
parametrisable \textit{trigger table} \citep{qr:triggercard}. When enabled, this
trigger is applied on analysis-object data. In a real experiment, the online
selection is often divided into several steps (or \textit{levels}). 
corresponding to the different trigger levels.
First-level triggers are fast and simple but based only on partial data as not
all detector front-ends are readable within the decision latency. 
Higher level triggers are more complex, of finer-but-not-final quality and
based on full detector data. 

Real triggers are thus intrinsically based on reconstructed data with a worse
resolution than final analysis information. On the contrary, the same
information is used in \textit{Delphes} for the trigger emulation and for final
analyses.

\section{\label{sec:vfd}Very forward detector simulation}

\begin{figure}[!ht]
\begin{center}
\includegraphics[width=\columnwidth]{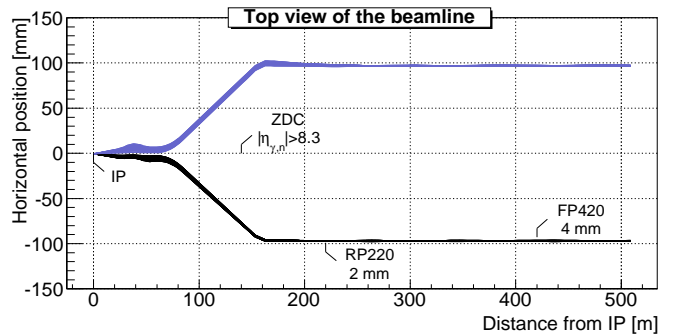}
\caption{Default location of the very forward detectors, including
\textsc{ZDC}, \textsc{RP220} and \textsc{FP420} in the \textsc{LHC} beamline.
Incoming (beam 1, red) and outgoing (beam 2, black) beams on one side of the
fifth interaction point (\textsc{IP5}, $s=0~\textrm{m}$ on the plot).
The Zero Degree Calorimeter is located in perfect alignment with the beamline
axis at the interaction point, at $140~\textrm{m}$, where the beam paths are
well separated. The forward taggers are near-beam detectors located at
$220~\textrm{m}$ and $420~\textrm{m}$. Beamline simulation with
\textit{Hector}~\citep{bib:hector}. All very forward detectors are located
symmetrically around the interaction point. }
\label{fig:fdets}
\end{center}
\end{figure}

Collider experiments often have additional instrumentation along the beamline.
These extend the $\eta$ coverage to higher values, for the detection of very
forward final-state particles. In \textit{Delphes}, Zero Degree Calorimeters,
roman pots and forward taggers have been implemented (Fig.~\ref{fig:fdets}),
similarly as for CMS and ATLAS collaborations~\citep{bib:cmsjetresolution,
bib:ATLASresolution}.

\begin{table}[t]
\begin{center}
\caption{Default parameters for the forward detectors: distance from the
interaction point and detector acceptance. The \textsc{LHC} beamline is assumed
around the fifth \textsc{LHC} interaction point (\textsc{IP}). For the
\textsc{ZDC}, the acceptance depends only on the pseudorapidity $\eta$ of the
particle, which should be neutral and stable.
It is expressed in terms of the particle energy ($E$).
All detectors are located on both sides of the interaction point.
\vspace{0.5cm}}
\begin{tabular}{llcl}
\hline
Detector & Distance & Acceptance & \\ \hline
\textsc{ZDC}   & $\pm 140$ m & $|\eta|> 8.3$       & for $n$ and $\gamma$\\
\textsc{RP220} & $\pm 220$ m & $E \in [6100 ; 6880]$ (GeV) & at $2~\textrm{mm}$\\
\textsc{FP420} & $\pm 420$ m & $E \in [6880 ; 6980]$ (GeV) & at $4~\textrm{mm}$\\
\hline
\end{tabular}
\label{tab:fdetacceptance}
\end{center}
\end{table}

\subsection{Zero Degree Calorimeters}

In direct sight of the interaction point, on both sides of the central
detector, the Zero Degree Calorimeters (\textsc{ZDC}s) are located at zero
angle, i.e.\ are aligned with the beamline axis at the interaction point. They
are placed beyond the point where the paths of incoming and outgoing beams
separate. These allow the measurement of stable neutral particles ($\gamma$ and
$n$) coming from the interaction point, with large pseudorapidities (e.g.\
$|\eta_{\textrm{n,}\gamma}| > 8.3$ in \textsc{ATLAS} and \textsc{CMS}).

The trajectory of the neutrals observed in the \textsc{ZDC}s is a straight
line, while charged particles are deflected away from their acceptance window by
the powerful magnets located in front of them. The fact that additional charged
particles may enter the \textsc{ZDC} acceptance is neglected in the current
versions of \textit{Delphes}.

The \textsc{ZDC}s have the ability to measure the time-of-flight of the particle.
This corresponds to the delay $t$ after which the particle is observed in the
detector, with respect to the bunch crossing reference time at the interaction
point ($t_0$): 
\begin{equation}
 t = t_0 + \frac{1}{v} \times \Big( \frac{s-z}{\cos \theta}\Big) \approx \frac{1}{c} \times (s-z),
\end{equation}
where $t_0$ is thus the true time coordinate of the vertex from which the
particle originates, $v$ the particle velocity, $s$ is the \textsc{ZDC} distance
to the interaction point, $z$ is the longitudinal coordinate of the vertex,
$\theta$ is the particle emission angle. It is assumed that the neutral particle
observed in the \textsc{ZDC} is highly relativistic and very forward.
For the time-of-flight measurement, a Gaussian smearing can be applied according
to the detector resolution
(Tab.~\ref{tab:defResolZdc})~\citep{qr:resolutionterms}.

The \textsc{ZDC}s are composed of an electromagnetic and a hadronic sections,
for the measurement of photons and neutrons, respectively. The energy of the
observed neutral is smeared according to Eq.~\ref{eq:caloresolution} and the
corresponding section resolutions (Tab.~\ref{tab:defResolZdc}). The \textsc{ZDC}
hits do not enter in the calorimeter cell list used for reconstruction of jets
and missing transverse energy.

\begin{table}[!h]
\begin{center}
\caption{Default values for the resolution of the zero degree calorimeters.
Resolution on energy measurement is parametrised by the \textit{stochastic}
($S$), \textit{noise} ($N$) and \textit{constant} ($C$) terms
(Eq.~\ref{eq:caloresolution})~\citep{qr:resolutionterms}. The time-of-flight is
smeared according to a Gaussian function.
\vspace{0.5cm}}
\begin{tabular}[!h]{llcc}
\hline
 \multicolumn{3}{l}{\textsc{ZDC}, electromagnetic part} & hadronic part \\
        & $S$ (GeV$^{1/2}$) 	& $0.7$ & $1.38$\\
        & $N$ (GeV)		& $0$ 	& $0$   \\
        & $C$ 			& $0.08$& $0.13$ \\
  \multicolumn{4}{l}{\textsc{ZDC}, timing resolution} \\
        & $\sigma_t$ (s) & $0$ & \\
\hline
\end{tabular}
\label{tab:defResolZdc}
\end{center}
\end{table}

The reconstructed ZDC hits correspond to neutral particles with a lifetime long
enough to reach these detectors (default: $c \tau \geq 140~\textrm{m}$) and very
large pseudorapidities (default: $|\eta|>8.3$). 
Photons and neutrons are identified if their energy overpasses a given threshold
(def. $E_\gamma \leq 20$~GeV and $E_n \leq 50$~GeV)~\citep{qr:fwdneutrals}.

\subsection{Forward taggers}

Forward taggers (called here \textsc{RP220}, for ``roman pots at
$220~\textrm{m}$'' and \textsc{FP420} for ``forward proton taggers at
$420~\textrm{m}$'', as at the \textsc{LHC}) are meant for the measurement of
particles following very closely the beam path. Such devices, also used at
\textsc{HERA} and Tevatron, are located very far away from the interaction point
(further than $150$~m in the \textsc{LHC} case).

To be able to reach these detectors, particles must have a charge identical to
the beam particles, and a momentum very close to the nominal value of the beam
particules. These taggers are near-beam detectors located a few millimetres from
the true beam trajectory and this distance defines their acceptance
(Tab.~\ref{tab:fdetacceptance}). For instance, roman pots at $220~\textrm{m}$
from the  \textsc{IP} and $2~\textrm{mm}$ from the beam will detect all forward
protons with an energy between $120$ and $900~\textrm{GeV}$~\citep{bib:hector}.
In \textit{Delphes}, extra hits coming from the beam-gas events or
secondary particles hitting the beampipe in front of the detectors are not taken
into account.

While neutral particles propagate along a straight line to the \textsc{ZDC}, a
dedicated simulation of the transport of charged particles is needed for
\textsc{RP220} and \textsc{FP420}. This fast simulation uses the \textit{Hector}
software~\citep{bib:hector}, which includes the chromaticity effects and the
geometrical aperture of the beamline elements of any arbitrary collider.

Forward taggers are able to measure the hit positions ($x,y$) and angles
($\theta_x,\theta_y$) in the transverse plane at the location of the detector
($s$ meters away from the \textsc{IP}), as well as the
time-of-flight\footnote{It is worth noting that for both \textsc{CMS} and
\textsc{ATLAS} experiments, the taggers located at $220$~m are not able to
measure the time-of-flight, contrary to \textsc{FP420} detectors.} ($t$). Out of
these the particle energy ($E$) and the momentum transfer it underwent during
the interaction ($q^2$) can be reconstructed at the analysis level (it is not
implemented in the current versions of \textit{Delphes}). The time-of-flight
measurement can be smeared with a Gaussian distribution (default value
$\sigma_t = 0~\textrm{s}$)~\citep{qr:protontaggers}.

\section{Validation}

\textit{Delphes} performs a fast simulation of a collider experiment.
Its performances in terms of computing time and data size are directly
proportional to the number of simulated events and on the considered physics
process. As an example, $10,000$ $pp \rightarrow t \bar t X$ events are
processed in $110~\textrm{s}$ on a regular laptop and use less than
$250~\textrm{MB}$ of disk space.
The quality and validity of the output are assessed by comparing the
resolutions on the reconstructed data to the expectations of both
\textsc{CMS}~\citep{bib:cmsjetresolution} and
\textsc{ATLAS}~\citep{bib:ATLASresolution} detectors.

Electrons and muons resolutions in \textit{Delphes} match by construction the
experiment designs, as the Gaussian smearing of their kinematics properties is
defined according to the detector specifications. Similarly, the $b$-tagging
efficiency (for real $b$-jets) and misidentification rates (for fake $b$-jets)
are taken directly from the expected values of the experiment. Unlike these
simple objects, jets and missing transverse energy should be carefully
cross-checked.

\subsection{Jet resolution}
 
The majority of interesting processes at the \textsc{LHC} contain jets in the
final state. The jet resolution obtained using \textit{Delphes} is therefore a
crucial point for its validation, both for \textsc{CMS}- and \textsc{ATLAS}-like
detectors. This validation is based on $pp \rightarrow gg$ events produced with
MadGraph/MadEvent and hadronised
using \textit{Pythia}~\citep{bib:mgme,bib:pythia}.

For a \textsc{CMS}-like detector, a similar procedure as the one explained in
published results is applied here. The events were arranged in $14$ bins of
gluon transverse momentum $\hat{p}_T$. In each $\hat{p}_T$ bin, every jet in
\textit{Delphes} is matched to the closest jet of generator-level particles,
using the spatial separation between the two jet axes
\begin{equation}
\Delta R = \sqrt{ \big(\eta^\textrm{rec} - \eta^\textrm{MC} \big)^2 + 
\big(\phi^\textrm{rec} - \phi^\textrm{MC} \big)^2}<0.25. 
\end{equation}
The jets made of generator-level particles, here referred as \textit{MC jets},
are obtained by applying the algorithm to all particles considered as stable
after hadronisation. Jets produced by \textit{Delphes}
and satisfying the matching criterion are called hereafter \textit{reconstructed
jets}. All jets are computed with the clustering algorithm (JetCLU) with a cone
radius $R$ of $0.7$.

The ratio of the transverse energies of every reconstructed jet
$E_T^\textrm{rec}$ to its corresponding \textsc{MC} jet $E_T^\textrm{MC}$ is
calculated in each $\hat{p}_T$ bin. The $E_T^\textrm{rec}/E_T^\textrm{MC}$
histogram is fitted with a Gaussian distribution in the interval \mbox{$\pm
2$~\textsc{RMS}} centred around the mean value. The resolution in each
$\hat{p}_T$ bin is obtained by the fit mean $\langle x \rangle$ and variance
$\sigma^2(x)$:
\begin{equation}
\frac{\sigma \Big (\frac{E_T^\textrm{rec}}{E_T^\textrm{MC}} \Big)_\textrm{fit}}{
\Big \langle \frac{E_T^\textrm{rec}}{E_T^\textrm{MC}} \Big
\rangle_\textrm{fit}}~
\Big( \hat{p}_T(i) \Big)\textrm{, for all }i.
\end{equation}
 
\begin{figure}[!ht]
\begin{center}
\includegraphics[width=\columnwidth]{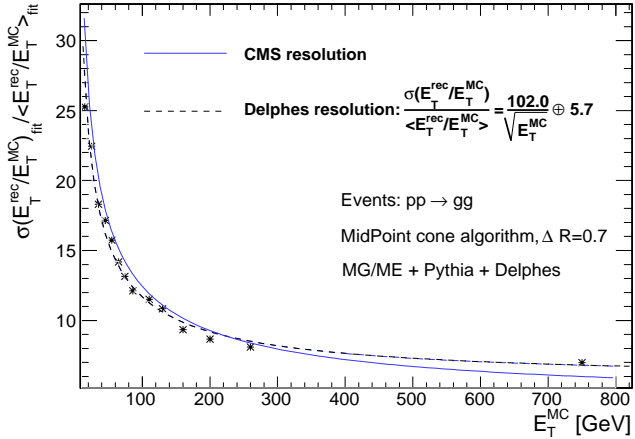}
\caption{Resolution of the transverse energy of reconstructed jets
$E_T^\textrm{rec}$ as a function of the transverse energy of the closest jet of
generator-level particles $E_T^\textrm{MC}$, in a \textsc{CMS}-like detector.
The jets events are reconstructed with the JetCLU clustering algorithm with a
cone radius of $0.7$. The maximum separation between the reconstructed and
\textsc{MC}-jets is $\Delta R= 0.25$. Dotted line is the fit result for
comparison to the \textsc{CMS} resolution~\citep{bib:cmsjetresolution}, in blue.
The $pp \rightarrow gg$ dijet events have been generated with MadGraph/MadEvent
and hadronised with \textit{Pythia}.}
\label{fig:jetresolcms}
\end{center}
\end{figure}

The resulting jet resolution as a function of $E_T^\textrm{MC}$ is shown in
Fig.~\ref{fig:jetresolcms}. 
This distribution is fitted with a function of the following form:
\begin{equation}
\frac{a}{E_T^\textrm{MC}}\oplus \frac{b}{\sqrt{E_T^\textrm{MC}}}\oplus c,
\label{eq:fitresolution}
\end{equation}
where $a$, $b$ and $c$ are the fit parameters. 
It is then compared to the resolution published by the \textsc{CMS}
collaboration~\citep{bib:cmsjetresolution}. The resolution curves from
\textit{Delphes} and \textsc{CMS} are in good agreement.

Similarly, the jet resolution is evaluated for an \textsc{ATLAS}-like detector.
The $pp \rightarrow gg$ events are here arranged in $8$ adjacent bins in $p_T$.
A $k_T$ reconstruction algorithm with $R=0.6$ is chosen and the maximal matching
distance between the \textsc{MC}-jets and the reconstructed jets is set to
$\Delta R=0.2$. The relative energy resolution is evaluated in each bin by:
\begin{equation}
\frac{\sigma(E)}{E} = \sqrt{~~ \Bigg \langle ~\Bigg( \frac{E^\textrm{rec} -
E^\textrm{MC}}{E^\textrm{rec}} \Bigg)^2 ~ \Bigg \rangle ~ - ~ \Bigg \langle
\frac{E^\textrm{rec} - E^\textrm{MC}}{ E^\textrm{rec} } \Bigg \rangle^2}.
\end{equation}

\begin{figure}[!ht]
\begin{center}
\includegraphics[width=\columnwidth]{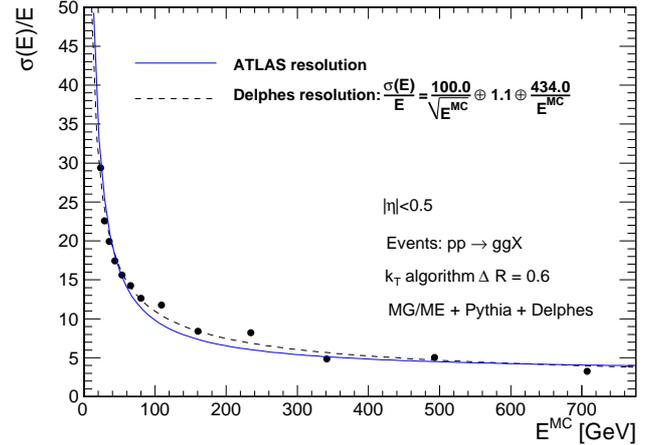}
\caption{Relative energy resolution of reconstructed jets as a function of the
energy of the closest jet of generator-level particles $E^\textrm{MC}$, in an
\textsc{ATLAS}-like detector. The jets are reconstructed with the $k_T$
algorithm with a radius $R=0.6$. The maximal matching distance between
\textsc{MC}- and reconstructed jets is $\Delta R=0.2$. Only central jets are
considered ($|\eta|<0.5$). Dotted line is the fit result for comparison to the
\textsc{ATLAS} resolution~\citep{bib:ATLASresolution}, in blue. The $pp
\rightarrow gg$ di-jet events have been generated with MadGraph/MadEvent and
hadronised with \textit{Pythia}.}
\label{fig:jetresolatlas}
\end{center}
\end{figure}

Figure~\ref{fig:jetresolatlas} shows a good agreement between the resolution
obtained with \textit{Delphes}, the result of the fit with
Equation~\ref{eq:fitresolution} and the corresponding curve provided by the
\textsc{ATLAS} collaboration~\citep{bib:ATLASresolution}.

\subsection{MET resolution}
 
All major detectors at hadron colliders have been designed to be as hermetic as
possible in order to detect the presence of one or more neutrinos and/or new
weakly interacting particles through apparent missing transverse energy.
The resolution of the $\overrightarrow{E_T}^\textrm{miss}$ variable, as
obtained with \textit{Delphes}, is then crucial.

The samples used to study the \textsc{MET} performance are identical to those
used for the jet validation. It is worth noting that the contribution to
$E_T^\textrm{miss}$ from muons is negligible in the studied sample.
The input samples are divided in five bins of scalar $E_T$ sums $(\Sigma E_T)$.
This sum, called \textit{total visible transverse energy}, is defined as the
scalar sum of transverse energy in all cells. The quality of the \textsc{MET}
reconstruction is checked via the resolution on its horizontal component
$E_x^\textrm{miss}$.

\begin{figure}[!ht]
\begin{center}
\includegraphics[width=\columnwidth]{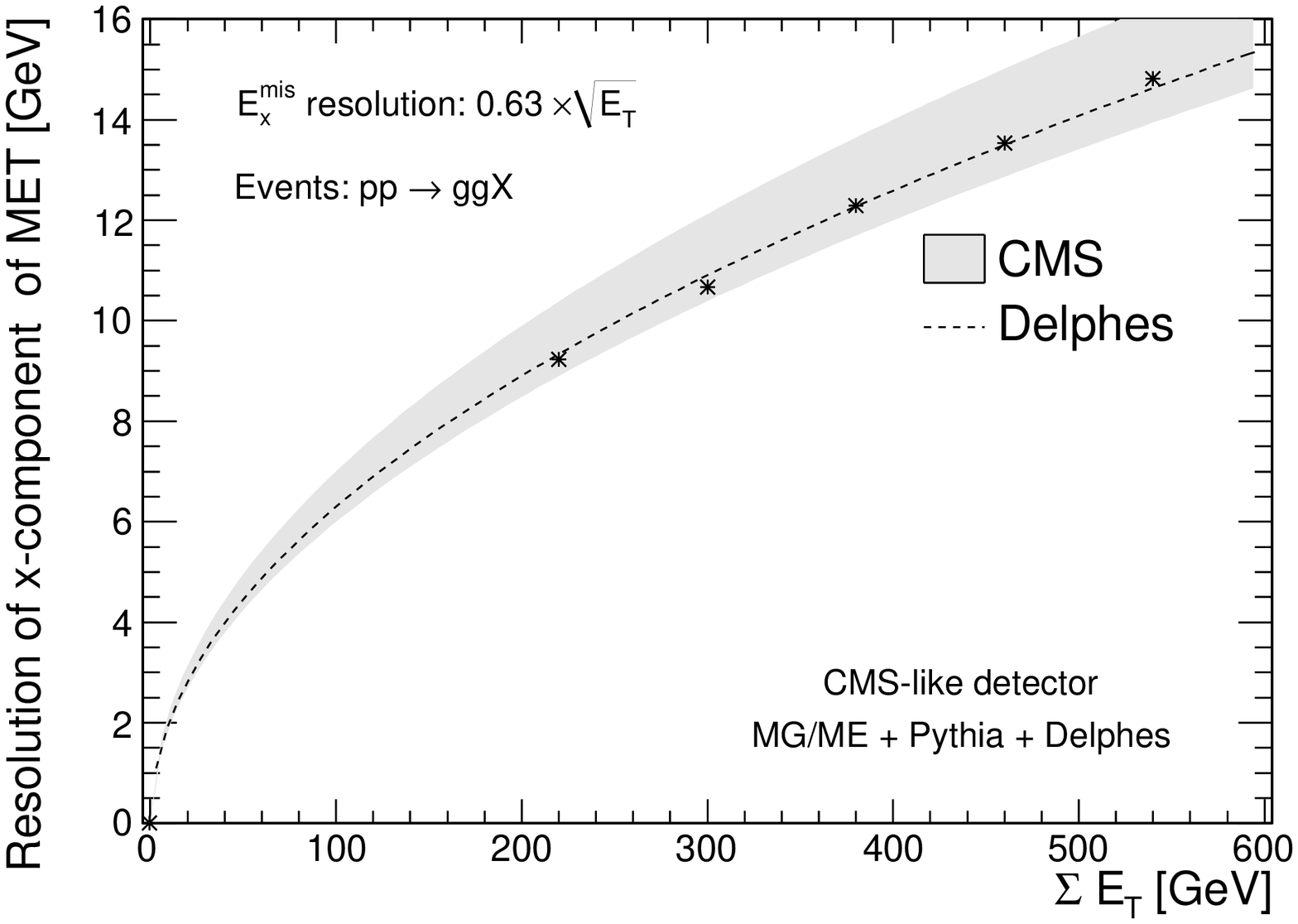}
\includegraphics[width=\columnwidth]{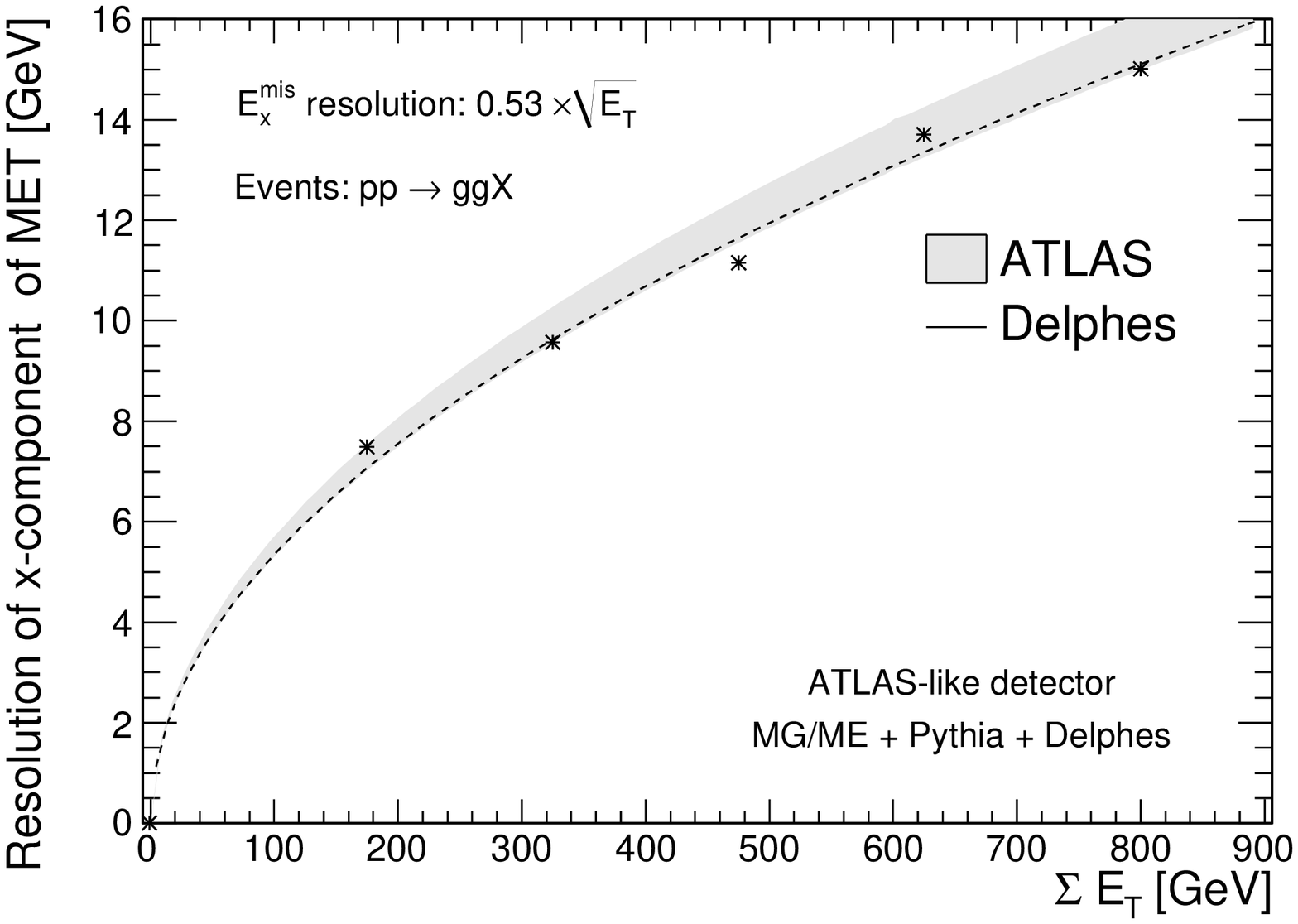}
\caption{$\sigma(E^\textrm{mis}_{x})$ as a function on the scalar sum of all
cells ($\Sigma E_T$) for $pp \rightarrow gg$ events, for a \textsc{CMS}-like
detector (top) and an \textsc{ATLAS}-like detector (bottom), for di-jet events
produced with MadGraph/MadEvent and hadronised with \textit{Pythia}.}
\label{fig:resolETmis}
\end{center}
\end{figure}

The $E_x^\textrm{miss}$ resolution is evaluated in the following way. The
distribution of the difference between $E_x^\textrm{miss}$ in \textit{Delphes}
and at generator-level is fitted with a Gaussian function in each $(\Sigma E_T)$
bin. The fit \textsc{RMS} gives the \textsc{MET} resolution in each bin.
The resulting value is presented in Fig.~\ref{fig:resolETmis} as a function of
the total visible transverse 
energy, for \textsc{CMS}- and \textsc{ATLAS}-like detectors.
 
The resolution $\sigma_x$ of the horizontal component of \textsc{MET} is
observed to behave like
\begin{equation}
\sigma_x = \alpha ~\sqrt{E_T}~~~(\mathrm{GeV}^{1/2}),
\end{equation}
where the $\alpha$ parameter depends on the resolution of the calorimeters. 

The \textsc{MET} resolution expected for the \textsc{CMS} detector for similar
events is $\sigma_x = (0.6-0.7) ~ \sqrt{E_T} ~ \mathrm{GeV}^{1/2}$ with no
pile-up (i.e. extra simultaneous $pp$ collision occurring at high-luminosity in
the same bunch crossing)~\citep{bib:cmsjetresolution}, which compares very well
with the $\alpha = 0.63$ obtained with \textit{Delphes}. Similarly, for an
\textsc{ATLAS}-like detector, a value of $0.53$ is obtained by \textit{Delphes}
for the $\alpha$ parameter, while the experiment expects it in the range $[0.53~
;~0.57]$~\citep{bib:ATLASresolution}.

\subsection{\texorpdfstring{$\tau$}{\texttau}-jet efficiency}
Table~\ref{tab:taurecoefficiency} lists the reconstruction efficiencies in
\textit{Delphes} for the hadronic $\tau$-jets from $H,Z \rightarrow \tau^+
\tau^-$. The mass of the Higgs boson is set successively to $140$ and
$300~\textrm{GeV}/c^2$. The inclusive gauge boson productions  ($pp \rightarrow
HX$ and $pp \rightarrow ZX$) are performed with MadGraph/MadEvent and the $\tau$
lepton decay and further hadronisation are handled by \textit{Pythia/Tauola}.
All reconstructed $\tau$-jets are $1-$prong, and follow the definition described
in section~\ref{btagging}, which is very close to an algorithm of the
\textsc{CMS} experiment~\citep{bib:cmstauresolution}. At last, corresponding
efficiencies published by the \textsc{CMS} and \textsc{ATLAS} experiments are
quoted for comparison. The level of agreement is satisfactory provided possible
differences due to the event generation chain and the detail of reconstruction
algorithms.

\begin{table}[!h]
\begin{center}
\caption{Reconstruction efficiencies of $\tau$-jets in $\tau^+ \tau^-$ decays
from $Z$ or $H$ bosons, in \textit{Delphes}, \textsc{CMS} and \textsc{ATLAS}
experiments~\citep{bib:cmstauresolution,bib:ATLASresolution}. Two scenarios for
the mass of the Higgs boson are investigated. Events generated with
MadGraph/MadEvent and hadronised with \textit{Pythia}. The decays of $\tau$
leptons is handled by the \textit{Tauola} version embedded in
\textit{Pythia}.\vspace{0.5cm}}
\begin{tabular}{lrlrl} 
\hline
				& \textsc{CMS}&Delphes & \textsc{ATLAS}&Delphes 
\\
$Z \rightarrow \tau^+ \tau^-$	& $38.2\%$ & $32.4\pm1.8\%$	& $33\%$ & $28.6\pm 1.9\%$ 		\\
$H(140) \rightarrow \tau^+ \tau^-$	& $36.3\%$ & $39.9\pm1.6\%$	& & $32.8\pm 1.8\%$ 		\\
$H(300) \rightarrow \tau^+ \tau^-$	& $47.3\%$ & $49.7\pm1.5\%$	& & $43.8\pm 1.6\%$ 		\\
\hline

\end{tabular}
\label{tab:taurecoefficiency}
\end{center}
\end{table}

\section{Visualisation}

When performing an analysis, a visualisation tool is useful to convey
information about the detector layout and the event topology in a simple way.
The \textit{Fast and Realistic OpenGL Displayer} \textsc{FROG}~\citep{bib:FROG}
has been interfaced in \textit{Delphes}, allowing an easy display of the defined
detector configuration~\citep{qr:frog}.
 
\begin{figure}[!ht]
\begin{center}
\includegraphics[width=\columnwidth]{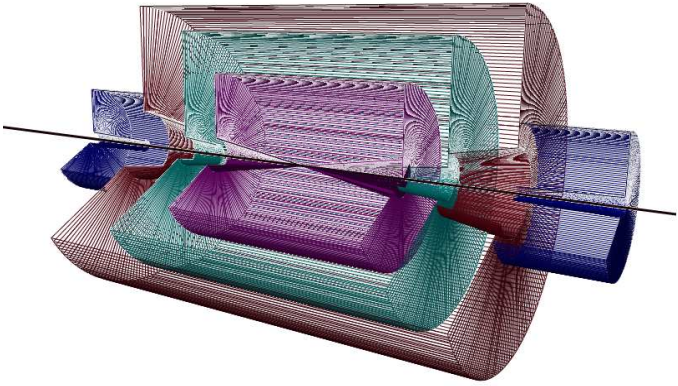}
\caption{Layout of the generic detector geometry assumed in \textit{Delphes}.
Open 3D-view of the detector with solid volumes. Same colour codes as for
Fig.~\ref{fig:GenDet3} are applied. Additional forward detectors are not
depicted.}
\label{fig:GenDet2}
\end{center}
\end{figure}

Two and three-dimensional representations of the detector configuration can be
used for communication purposes, as they clearly illustrate the geometric
coverage of the different detector subsystems. 
As an example, the generic detector geometry assumed in this paper is shown in
Fig.~\ref{fig:GenDet3} and~\ref{fig:GenDet2}. The extensions of the central
tracking system, the central calorimeters and both forward calorimeters are
visible. Note that only the geometrical coverage is depicted and that the
calorimeter segmentation is not taken into account in the drawing of the
detector. 
 
\begin{figure}[!ht]
\begin{center}
\includegraphics[width=0.9\columnwidth]{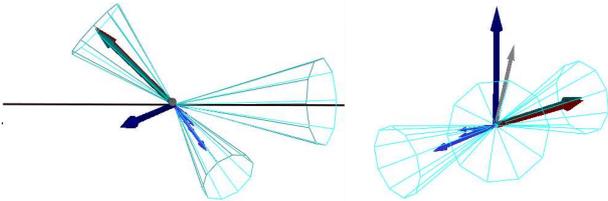}
\caption{Example of $pp(\gamma p \rightarrow Wt)pY$ event display in side (left)
and transverse (right) views, with $t \rightarrow Wb$. One
$W$ boson decays into a $\mu \nu_\mu$ pair and the second one into a $e \nu_e$
pair. The surviving proton leaves a forward hemisphere with no hadronic
activity. The isolated muon is shown as the dark blue vector. Around the
electron, in red, is reconstructed a fake $\tau$-jet (blue cone surrounding a
green arrow). The reconstructed missing energy is visible in grey. }
\label{fig:wt}
\end{center}
\end{figure}

Deeper understanding of interesting physics processes is possible by displaying
the events themselves. The visibility of each set of objects ($e^\pm$,
$\mu^\pm$, $\tau^\pm$, jets, transverse missing energy) is enhanced by a colour
coding. Moreover, kinematics information of each object is visible by a simple
mouse action. As an illustration, an associated photoproduction of a $W$ boson
and a $t$ quark~\citep{bib:wtphotoproduction} is shown in Fig.~\ref{fig:wt}. 


\begin{figure}[!ht]
\begin{center}
\includegraphics[width=0.6\columnwidth]{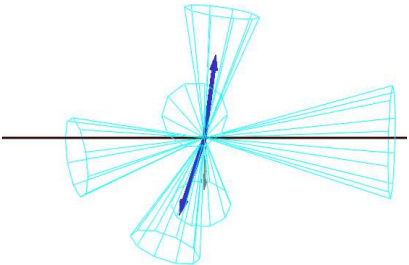}
\caption{Example of inclusive gluon pair production $pp \rightarrow ggX$. Many
jets are present in the event, in particular along the beam axis (black line).}
\label{fig:gg}
\end{center}
\end{figure}

For comparison, Fig.~\ref{fig:gg} depicts an inclusive gluon pair production
$pp \rightarrow ggX$. The event final state contains more jets, in particular
along the beam axis, which is expected as the interacting protons are destroyed
by the collision. 

\section{Conclusion and perspectives}

We have described here the major features of the \textit{Delphes} framework,
introduced for the fast simulation of a collider experiment. This framework is a
tool meant for feasibility studies in phenomenology, gauging the observability
of model predictions in collider experiments.

\textit{Delphes} takes as an input the output of event-generators and yields
analysis-object data in the form of \texttt{TTree} in a \texttt{*.root} file.
The simulation includes central and forward detectors to produce realistic
observables using standard reconstruction algorithms. 
Moreover, the framework allows trigger emulation and 3D event visualisation.

\textit{Delphes} has been developed using the parameters of the \textsc{CMS}
experiment but can be easily extended to \textsc{ATLAS} and other
non-\textsc{LHC} experiments, as at Tevatron or at the \textsc{ILC}. Further
developments include a more flexible design for the subdetector assembly, a
better $b$-tag description and possibly the implementation of an event mixing
module for pile-up event simulation. This framework has already been used for
several analyses~\citep{bib:wtphotoproduction, bib:papierquisortirajamais,
bib:papiersimon}, in particular in photon-induced interactions at the
\textsc{LHC}.

\section*{Acknowledgements}
\addcontentsline{toc}{section}{Acknowledgements}
The authors would like to thank Jer\^ome de Favereau, Christophe Delaere, Muriel Vander Donckt and David d'Enterria for useful discussions and comments, and Loic Quertenmont for support in interfacing \textsc{FROG}. We are also really grateful to Alice Dechambre and Simon de Visscher for being beta testers of the complete package.
Part of this work was supported by the Belgian Federal Office for Scientific, Technical and Cultural Affairs through the Interuniversity Attraction Pole P6/11.

\renewcommand\refname{Internal code references}

\onecolumn
\appendix

\section{User manual}
 
The available \texttt{C++}-code is compressed in a zipped tar file which contains everything needed to run the \textit{Delphes} package, assuming a running \textsc{ROOT} installation. The package includes \texttt{ExRootAnalysis}~\citep{bib:ExRootAnalysis}, \textit{Hector}~\citep{bib:hector}, FastJet~\citep{bib:FASTJET}, and \textsc{FROG}~\citep{bib:FROG}, as well as the conversion codes to read standard \mbox{StdHEP} input files (\texttt{mcfio} and \texttt{stdhep})~\citep{bib:mcfio} and HepMC~\citep{bib:hepmc}.
In order to visualise the events with the \textsc{FROG} software, a few additional external libraries may be required, as explained in \href{http://projects.hepforge.org/FROG/}{http://projects.hepforge.org/FROG/}.
 
\subsection{Getting started}
 
In order to run \textit{Delphes} on your system, first download its sources and compile them:\\
\begin{quote}\texttt{wget http://www.fynu.ucl.ac.be/users/s.ovyn/Delphes/files/Delphes\_V\_*.tar.gz}\end{quote}
Replace the \texttt{*} symbol by the proper version number. Always refer to the download page on the \textit{Delphes} website \href{http://www.fynu.ucl.ac.be/users/s.ovyn/Delphes/download.html}{http://www.fynu.ucl.ac.be/users/s.ovyn/Delphes/download.html}. Current version of Delphes for this manual is V 1.8 (July 2009).

\begin{quote}
\begin{verbatim}
me@mylaptop:~$ tar -xvf Delphes_V_*.tar.gz
me@mylaptop:~$ cd Delphes_V_*.*
me@mylaptop:~$ ./genMakefile.tcl > Makefile
me@mylaptop:~$ make
\end{verbatim}
\end{quote}
Due to the large number of external utilities, the number of printed lines during the compilation can be high. The user should not pay attention to possible warning messages, which are due to the external packages used by \textit{Delphes}. When compilation is completed, the following message is printed:
\begin{quote}
\begin{verbatim}
me@mylaptop:~$ Delphes has been compiled
me@mylaptop:~$ Ready to run
\end{verbatim}
\end{quote}

\subsection{Running \textit{Delphes} on your events}
 
In this sub-appendix, we will explain how to use \textit{Delphes} to perform a fast simulation of a general-purpose detector on your event files. The first step to use \textit{Delphes} is to create the list of input event files (e.g.\ {\verb inputlist.list }). It is important to notice that all the files comprised in the list file should have the same of extension (\texttt{*.hep}, \texttt{*.lhe}, \texttt{*.hepmc} or \texttt{*.root}). In the simplest way to run \textit{Delphes}, you need this input file and you need to specify the name of the output file that will contain the generator-level data (\texttt{GEN} tree), the analysis data objects after reconstruction (\texttt{Analysis} tree), and the results of the trigger emulation (\texttt{Trigger} tree).
 
\begin{quote}
\begin{verbatim}
me@mylaptop:~$ ./Delphes inputlist.list OutputRootFileName.root
\end{verbatim}
\end{quote}
 
\subsubsection{Setting up the configuration}
 
The program is driven by two datacards (default cards are {\verb data/DetectorCard.dat } and {\verb data/TriggerCard.dat }) which allow the user to choose among a large spectrum of running conditions. Please note that if the user does not provide these datacards, the running will be done using the default parameters defined in the constructor of the class \texttt{RESOLution} (see next). If you choose a different detector or running configuration, you will need to edit the datacards accordingly. Detector and trigger cards are provided in the \texttt{data/} subdirectory for the \textsc{CMS} and \textsc{ATLAS} experiments.
 
\begin{enumerate}
\item{\bf The detector card }
It contains all pieces of information needed to run \textit{Delphes}:
\begin{itemize}
 \item detector parameters, including calorimeter and tracking coverage and resolutions, transverse energy thresholds for object reconstruction and jet algorithm parameters.
 \item six flags ({\verb FLAG_bfield }, {\verb FLAG_vfd }, {\verb FLAG_RP }, {\verb FLAG_trigger }, {\verb FLAG_FROG } and {\verb FLAG_LHCO }), should be set in order to configure the magnetic field propagation, the very forward detectors simulation, the use of very forward taggers, the trigger selection, the preparation for \textsc{FROG} display and the creation of an output file in \texttt{*.LHCO} text format (respectively).
 \end{itemize}
 
If no datacard is provided by the user, the default smearing and running parameters are used (corresponding to tables~\ref{tab:defEta},~\ref{tab:defResol}).\\
Definition of the sub-detector extensions:
\begin{quote}
\begin{verbatim}
CEN_max_tracker    2.5     // Maximum tracker coverage
CEN_max_calo_cen   1.7     // central calorimeter coverage
CEN_max_calo_ec    3.0     // calorimeter endcap coverage
CEN_max_calo_fwd   5.0     // forward calorimeter pseudorapidity coverage
CEN_max_mu         2.4     // muon chambers pseudorapidity coverage
\end{verbatim}
\end{quote}
Definition of the sub-detector resolutions:
\begin{quote}
\begin{verbatim}
# Energy resolution for electron/photon in central/endcap/fwd/zdc calos
# \sigma/E = C + N/E + S/\sqrt{E}, E in GeV
ELG_Scen          0.05     // S term for central ECAL
ELG_Ncen          0.25     // N term 
ELG_Ccen          0.005    // C term 
ELG_Sec           0.05     // S term for ECAL endcap
ELG_Nec           0.25     // N term
ELG_Cec           0.005    // C term
ELG_Sfwd          2.084    // S term for FCAL
ELG_Nfwd          0.       // N term 
ELG_Cfwd          0.107    // C term 
ELG_Szdc          0.70     // S term for ZDC
ELG_Nzdc          0.       // N term 
ELG_Czdc          0.08     // C term 

# Energy resolution for hadrons in central/endcap/fwd/zdc calos
# \sigma/E = C + N/E + S/\sqrt{E}, E in GeV
HAD_Scen          1.5      // S term for central HCAL
HAD_Ncen          0.       // N term 
HAD_Ccen          0.05     // C term
HAD_Sec           1.5      // S term for HCAL endcap
HAD_Nec           0.       // N term
HAD_Cec           0.05     // C term 
HAD_Sfwd          2.7      // S term for FCAL
HAD_Nfwd          0.       // N term 
HAD_Cfwd          0.13     // C term 
HAD_Szdc          1.38     // S term for ZDC
HAD_Nzdc          0.       // N term 
HAD_Czdc          0.13     // C term

# Time resolution for ZDC/RP220/RP420
ZDC_T_resolution   0       // in s
RP220_T_resolution 0       // in s
RP420_T_resolution 0       // in s
 
# Muon smearing
MU_SmearPt        0.01     // transverse momentum Pt in GeV/c
 
# Tracking efficiencies
TRACK_ptmin       0.9      // minimal pT
TRACK_eff          90      // efficiency associated to the tracking (%)
\end{verbatim}
\end{quote}
Definitions related to the calorimetric cells:
\begin{quote}
\begin{verbatim}
# Calorimetric towers
TOWER_number         40
TOWER_eta_edges 0. 0.087 0.174 0.261 0.348 0.435 0.522 0.609 0.696 0.783
               0.870 0.957 1.044 1.131 1.218 1.305 1.392 1.479 1.566 1.653
               1.740 1.830 1.930 2.043 2.172 2.322 2.500 2.650 2.868 2.950
               3.125 3.300 3.475 3.650 3.825 4.000 4.175 4.350 4.525 4.700
               5.000
 
TOWER_dphi 5 5 5 5 5 5 5 5 5 5 5 5 5 5 5 5 5 5 5 10
           10 10 10 10 10 10 10 10 10 10 10 10 10 10 10 10 10 10 20 20
\end{verbatim}
\end{quote}
\texttt{TOWER\_eta\_edges} is the list of the edges in $\eta$ of all cells, in the $\eta>0$ hemisphere (the detector is supposed to be symmetric with respect to the $\eta=0$ plane, as well as around the $z$-axis). Starts with the lower edge of the most central tower (default: $\eta = 0$) and ends with the higher edge of the most forward tower.
\texttt{TOWER\_dphi} lists the tower size in $\phi$ (in degree), assuming that all cells are similar in $\phi$ for a given $\eta$.\\
Thresholds applied for storing the reconstructed objects in the final collections:
\begin{quote}
\begin{verbatim}
# Thresholds for reconstructed objects, in GeV/c
PTCUT_elec       10.0
PTCUT_muon       10.0
PTCUT_jet        20.0
PTCUT_gamma      10.0
PTCUT_taujet     10.0

# Thresholds for reconstructed objects in ZDC, E in GeV
ZDC_gamma_E      20
ZDC_n_E          50
\end{verbatim}
\end{quote}
Definitions of variables related to the charged lepton isolation:
\begin{quote}
\begin{verbatim}
# Charged lepton isolation. Pt and Et in GeV
ISOL_PT          2.0  //minimal pt of tracks for isolation criteria
ISOL_Cone        0.5  //Cone  for isolation criteria
ISOL_Calo_Cone   0.4  //Cone for calorimetric isolation
ISOL_Calo_ET     2.0  //minimal tower E_T for isolation criteria. 1E99 means "off"
ISOL_Calo_Grid   3    //Grid size (N x N) for calorimetric isolation
\end{verbatim}
\end{quote}
Definitions of variables related to the jet reconstruction:
\begin{quote}
\begin{verbatim}
# General jet variable
JET_coneradius   0.7  // generic jet radius 
JET_jetalgo      1    // 1 for Cone algorithm, 
                      // 2 for MidPoint algorithm, 
                      // 3 for SIScone algorithm, 
                      // 4 for kt algorithm
                      // 5 for Cambridge/Aachen algorithm
                      // 6 for anti-kt algorithm
JET_seed         1.0  // minimum seed to start jet reconstruction, in GeV
JET_Eflow        1    // Energy flow: perfect energy assumed in the tracker coverage.
                      // 1 is 'on' ; 0 is 'off'

# Tagging definition 
BTAG_b           40    // b-tag efficiency (%)
BTAG_mistag_c    10    // mistagging (%)
BTAG_mistag_l    1     // mistagging (%)
\end{verbatim}
\end{quote}
Switches for options
\begin{quote}
\begin{verbatim}
# FLAGS
FLAG_bfield      1     //1 to run the bfield propagation else 0
FLAG_vfd         1     //1 to run the very forward detectors else 0
FLAG_RP          1     //1 to run the very forward detectors else 0
FLAG_trigger     1     //1 to run the trigger selection else 0
FLAG_FROG        1     //1 to run the FROG event display
FLAG_LHCO        1     //1 to run the LHCO
\end{verbatim}
\end{quote}
Parameters for the magnetic field simulation:
\begin{quote}
\begin{verbatim}
# In case BField propagation allowed
TRACK_radius      129   // radius of the BField coverage, in cm
TRACK_length      300   // length of the BField coverage, in cm
TRACK_bfield_x    0     // X component of the BField, in T
TRACK_bfield_y    0     // Y component of the BField, in T
TRACK_bfield_z    3.8   // Z component of the BField, in T
\end{verbatim}
\end{quote}
Parameters related to the very forward detectors
\begin{quote}
\begin{verbatim}
# Very forward detector extension, in pseudorapidity
# if allowed
VFD_min_zdc       8.3   // Zero-Degree neutral Calorimeter
VFD_s_zdc         140   // distance of the ZDC, from the IP, in [m]

#\textit{Hector} parameters
RP_220_s          220     // distance of the RP to the IP, in meters
RP_220_x          0.002   // distance of the RP to the beam, in meters
RP_420_s          420     // distance of the RP to the IP, in meters
RP_420_x          0.004   // distance of the RP to the beam, in meters
RP_beam1Card      data/LHCB1IR5_v6.500.tfs // beam optics file, beam 1
RP_beam2Card      data/LHCB2IR5_v6.500.tfs // beam optics file, beam 2
RP_IP_name        IP5     // tag for IP in \textit{Hector} ; 'IP1' for ATLAS
RP_offsetEl_x     0.097   // horizontal separation between both beam, in meters
RP_offsetEl_y     0       // vertical separation between both beam, in meters
RP_offsetEl_s     120     // distance of beam separation point, from IP
RP_cross_x        -500    // IP offset in horizontal plane, in micrometers
RP_cross_y        0       // IP offset in vertical plane, in micrometers
RP_cross_ang_x    142.5   // half-crossing angle in horizontal plane, in microrad
RP_cross_ang_y    0       // half-crossing angle in vertical plane, in microrad
\end{verbatim}
\end{quote}
Others parameters:
\begin{quote}
\begin{verbatim}
# In case FROG event display allowed
NEvents_FROG      100
# Number of events to process
NEvents           -1                    // -1 means 'all'

# input PDG tables
PdgTableFilename  data/particle.tbl     // table with particle pid,mass,charge,...
\end{verbatim}
\end{quote}

In general, energies, momenta and masses are expressed in GeV, GeV$/c$, GeV$/c^2$ respectively, and  magnetic fields in T.
Geometrical extension are often referred in terms of pseudorapidity $\eta$, as the detectors are supposed to be symmetric in $\phi$. From version 1.8 onwards, the number of events to run is also be included in the detector card (\texttt{NEvents}). For version 1.7 and earlier, the parameters related to the calorimeter endcaps (\texttt{CEN\_max\_calo\_ec}, \texttt{ELG\_Sec}, \texttt{ELG\_Nec}, \texttt{ELG\_Cec}, \texttt{HAD\_Sec}, \texttt{HAD\_Nec} and \texttt{HAD\_Cec}) did not exist in the detector cards; in addition, some other variables had different names (\texttt{HAD\_Scen} was \texttt{HAD\_Sfcal}, \texttt{HAD\_Ncen} was \texttt{HAD\_Nfcal}, \texttt{HAD\_Ccen} was \texttt{HAD\_Cfcal}, \texttt{HAD\_Sfwd} was \texttt{HAD\_Shf}, \texttt{HAD\_Nfwd} was \texttt{HAD\_Nhf}, \texttt{HAD\_Cfwd} was \texttt{HAD\_Chf}). However, these cards are still completely compatible with new versions of \textit{Delphes}. In such a case, the calorimeter endcaps are simply assumed to be located at the edge of the central calorimeter volumes, with the same resolution values.
 
\item{\bf The trigger card }
 
This card contains the definitions of all trigger-bits. Cuts can be applied on the transverse momentum $p_T$ of electrons, muons, jets, $\tau$-jets, photons and the missing transverse energy. The following codes should be used so that \textit{Delphes} can correctly translate the input list of trigger-bits into selection algorithms:

\begin{quote}
\begin{tabular}{ll}
{\it Trigger code} & {\it Corresponding object}\\
{\verb ELEC_PT } & electron \\
{\verb IElec_PT } & isolated electron \\
{\verb MUON_PT } & muon \\
{\verb IMuon_PT } & isolated muon \\
{\verb JET_PT } & jet \\
{\verb TAU_PT } & $\tau$-jet \\
{\verb ETMIS_PT } & missing transverse energy \\
{\verb GAMMA_PT } & photon \\
{\verb Bjet_PT } & $b$-jet \\
\end{tabular}
\end{quote}
 
Each line in the trigger datacard is allocated to exactly one trigger-bit and starts with the name of the corresponding trigger. 
Logical combination of several conditions is also possible. If the trigger-bit requires the presence of multiple identical objects, the order of their $p_T$ thresholds is very important: they must be defined in \textit{decreasing} order. The transverse momentum $p_T$ is expressed in \mbox{GeV/$c$}. Finally, the different requirements on the objects must be separated by a {\verb && } flag. 
The default trigger card can be found in the data repository of \textit{Delphes} (\texttt{data/TriggerCard.dat}), as well as for both \textsc{CMS} and \textsc{ATLAS} experiments at the \textsc{LHC}.
An example of trigger table consistent with the previous rules is given here:
\begin{quote}
\begin{verbatim}   
SingleJet                   >> JET_PT: '200'
DoubleElec                  >> ELEC_PT: '20' && ELEC_PT: '10'   
SingleElec and Single Muon  >> ELEC_PT: '20' && MUON_PT: '15'
\end{verbatim}
\end{quote}
\end{enumerate}
 
\subsubsection{Running the code}
 
First, create the detector and trigger cards (\texttt{data/DetectorCard.dat} and \texttt{data/TriggerCard.dat}). \\
Then, create a text file containing the list of input files that will be used by \textit{Delphes} (with extension \texttt{*.lhe}, \texttt{*.hepmc}, \texttt{*.root} or \texttt{*.hep}).
To run the code, type the following command (in one line)
\begin{quote}
\begin{verbatim}
me@mylaptop:~$ ./Delphes inputlist.list OutputRootFileName.root
                         data/DetectorCard.dat data/TriggerCard.dat
\end{verbatim}
\end{quote}
As a reminder, typing the \texttt{./Delphes} command simply displays the correct usage:

\begin{quote}
\begin{verbatim}
me@mylaptop:~$ ./Delphes
 Usage: ./Delphes input_file output_file [detector_card] [trigger_card]
 input_list - list of files in Ntpl, StdHep, HepMC or LHEF format,
 output_file - output file.
 detector_card - Card containing resolution variables for detector simulation (optional)
 trigger_card - Card containing the trigger algorithms (optional)
\end{verbatim}
\end{quote}

\subsection{Getting the \textit{Delphes} information}
 
\subsubsection{Contents of the \textit{Delphes} ROOT trees}
 
The \textit{Delphes} output file (\texttt{*.root}) is subdivided into three \textit{trees}, corresponding to generator-level data, analysis-object data and trigger output. These \textit{trees} are structures that organise the output data into \textit{branches} containing data (or \textit{leaves}) related with each others, like the kinematics properties ($E$, $p_x$, $\eta$, $\ldots$) of a given particle.

Here is the exhaustive list of \textit{branches} availables in these \textit{trees}, together with their corresponding physical objet and \texttt{ExRootAnalysis} C++ class name:
\begin{quote}
\begin{tabular}{lll}
\textbf{GEN \texttt{Tree}} & &\\
~~~Particle & generator particles from \textsc{hepevt}     & {\verb GenParticle }\\
\multicolumn{3}{l}{}\\
\textbf{Trigger  \texttt{Tree}} & &\\
~~~TrigResult & Acceptance of different trigger-bits       & {\verb TRootTrigger }\\
\multicolumn{3}{l}{}\\
\textbf{Analysis \texttt{Tree}} & & \\
~~~Tracks     & Collection of tracks                       & {\verb TRootTracks }\\
~~~CaloTower  & Calorimetric cells                         & {\verb TRootCalo }\\
~~~Electron   & Collection of electrons                    & {\verb TRootElectron }\\
~~~Photon     & Collection of photons                      & {\verb TRootPhoton }\\
~~~Muon       & Collection of muons                        & {\verb TRootMuon }\\
~~~Jet        & Collection of jets                         & {\verb TRootJet }\\
~~~TauJet     & Collection of jets tagged as $\tau$-jets   & {\verb TRootTauJet }\\
~~~ETmis      & Transverse missing energy information      & {\verb TRootETmis }\\
~~~ZDChits    & Hits in the Zero Degree Calorimeters       & {\verb TRootZdcHits }\\
~~~RP220hits  & Hits in the first proton taggers           & {\verb TRootRomanPotHits }\\
~~~FP420hits  & Hits in the next proton taggers            & {\verb TRootRomanPotHits }\\
\end{tabular}
\end{quote}
The third column shows the names of the corresponding classes to be written in a \textsc{ROOT} tree.
The bin number in the unique leaf in the \texttt{trigger} tree (namely, \texttt{TrigResult.Accepted}) corresponds to the trigger number in the provided list. In addition, the result of the global trigger decision upon each event (i.e.\ the logical \texttt{OR} of all trigger conditions) is stored in the first bin (number 0) of this leaf.
In \texttt{Analysis} tree, all classes except \texttt{TRootTracks}, \texttt{TRootCalo}, \texttt{TRootTrigger}, \texttt{TRootETmis} and \texttt{TRootRomanPotHits} inherit from the class \texttt{TRootParticle} which includes the following data members (stored as \textit{leaves} in \textit{branches} of the \textit{trees}):
\begin{quote}
\begin{tabular}{ll}
\multicolumn{2}{l}{\textbf{Most common leaves}}\\
 \texttt{~~~float E;     }&\texttt{ // particle energy in GeV }\\
 \texttt{~~~float Px;    }&\texttt{ // particle momentum vector (x component) in GeV$/c$ }\\
 \texttt{~~~float Py;    }&\texttt{ // particle momentum vector (y component) in GeV$/c$ }\\
 \texttt{~~~float Pz;    }&\texttt{ // particle momentum vector (z component) in GeV$/c$ }\\
 \texttt{~~~float PT;    }&\texttt{ // particle transverse momentum in GeV$/c$ }\\
 \texttt{~~~float Eta;   }&\texttt{ // particle pseudorapidity   }\\
 \texttt{~~~float Phi;   }&\texttt{ // particle azimuthal angle in rad }\\
\end{tabular}
\end{quote}
 
In addition to their kinematics, some additional properties are available for specific objects: 
\begin{quote}
\begin{tabular}{ll}
\multicolumn{2}{l}{{\bf Leaves in the \texttt{Particle} branch (\texttt{GEN} tree)}} \\    
   \texttt{~~~int PID;      }&\texttt{ // particle HEP ID number }\\
   \texttt{~~~int Status;   }&\texttt{ // particle status }\\
   \texttt{~~~int M1;       }&\texttt{ // particle 1st mother }\\
   \texttt{~~~int M2;       }&\texttt{ // particle 2nd mother }\\
   \texttt{~~~int D1;       }&\texttt{ // particle 1st daughter }\\
   \texttt{~~~int D2;       }&\texttt{ // particle 2nd daughter }\\
   \texttt{~~~float Charge; }&\texttt{ // electrical charge in units of e}\\
   \texttt{~~~float T;      }&\texttt{ // particle vertex position (t component, in mm$/c$) }\\
   \texttt{~~~float X;      }&\texttt{ // particle vertex position (x component, in mm) }\\
   \texttt{~~~float Y;      }&\texttt{ // particle vertex position (y component, in mm) }\\
   \texttt{~~~float Z;      }&\texttt{ // particle vertex position (z component, in mm) }\\
   \texttt{~~~float M;      }&\texttt{ // particle mass in GeV$/c^2$}\\
\end{tabular}
\end{quote}
\begin{quote}
\begin{tabular}{ll}
\multicolumn{2}{l}{\textbf{Additional leaves in \texttt{Electron} and \texttt{Muon} branches} (\texttt{Analysis} tree)} \\
   \texttt{~~~int Charge }    &\texttt{ // particle Charge }\\
   \texttt{~~~bool IsolFlag } &\texttt{ // stores the result of the tracking isolation test }\\
   \texttt{~~~float IsolPt } &\texttt{  // sum of all track pt in isolation cone (GeV/c) }\\
   \texttt{~~~float EtaCalo } &\texttt{ // particle pseudorapidity when entering the calo }\\
   \texttt{~~~float PhiCalo } &\texttt{ // particle azimuthal angle in rad when entering the calo }\\
   \texttt{~~~float EHoverEE }&\texttt{ // hadronic energy over electromagnetic energy }\\
   \texttt{~~~float EtRatio } &\texttt{ // calo Et in NxN-cell grid around the muon over the muon Et }\\
\end{tabular}
\end{quote}
\begin{quote}
\begin{tabular}{ll}
\multicolumn{2}{l}{\textbf{Additional leaf in the \texttt{Jet} branch (\texttt{Analysis} tree)}}  \\
   \texttt{~~~bool Btag }  &\texttt{ // stores the result of the b-tagging }\\
   \texttt{~~~int NTracks }&\texttt{ // number of tracks associated to the jet }\\
   \texttt{~~~float EHoverEE }&\texttt{ // hadronic energy over electromagnetic energy }\\
\end{tabular}
\end{quote}
\begin{quote}
\begin{tabular}{ll}
\multicolumn{2}{l}{\textbf{Leaves in the \texttt{Tracks} branch (\texttt{Analysis} tree)}}\\
    \texttt{~~~float Eta }     &\texttt{ // pseudorapidity at the beginning of the track }\\
    \texttt{~~~float Phi }     &\texttt{ // azimuthal angle at the beginning of the track }\\
    \texttt{~~~float EtaOuter }&\texttt{ // pseudorapidity at the end of the track }\\
    \texttt{~~~float PhiOuter }&\texttt{ // azimuthal angle at the end of the track }\\
    \texttt{~~~float PT }      &\texttt{ // track transverse momentum in GeV$/c$ }\\ 
    \texttt{~~~float E }       &\texttt{ // track energy in GeV }\\
    \texttt{~~~float Px }      &\texttt{ // track momentum vector (x component) in GeV$/c$ }\\
    \texttt{~~~float Py }      &\texttt{ // track momentum vector (y component) in GeV$/c$ }\\
    \texttt{~~~float Pz }      &\texttt{ // track momentum vector (z component) in GeV$/c$ }\\
    \texttt{~~~float Charge }  &\texttt{ // track charge in units of $e$ }\\
\end{tabular}
\end{quote}
\begin{quote}
\begin{tabular}{ll}
\multicolumn{2}{l}{\textbf{Leaves in the \texttt{CaloTower} branch (\texttt{Analysis} tree)}}\\
    \texttt{~~~float Eta }     &\texttt{ // pseudorapidity of the cell }\\
    \texttt{~~~float Phi }     &\texttt{ // azimuthal angle of the cell in rad }\\
    \texttt{~~~float E }       &\texttt{ // cell energy in GeV }\\
    \texttt{~~~float E\_em }   &\texttt{ // electromagnetic component of the cell energy in GeV}\\
    \texttt{~~~float E\_had }  &\texttt{ // hadronic component of the cell energy in GeV}\\
    \texttt{~~~float ET }      &\texttt{ // cell transverse energy in GeV }\\
& \\
\multicolumn{2}{l}{\textbf{Leaves in the \texttt{ETmis} branch (\texttt{Analysis} tree)}}\\
    \texttt{~~~float Phi }     &\texttt{ // azimuthal angle of the transverse missing energy in rad }\\
    \texttt{~~~float ET }      &\texttt{ // transverse missing energy in GeV }\\
    \texttt{~~~float Px }      &\texttt{ // x component of the transverse missing energy in GeV }\\
    \texttt{~~~float Py }      &\texttt{ // y component of the transverse missing energy in GeV }\\
\end{tabular}
\end{quote}

The hits in very forward detector (\textsc{ZDC, RP220, FP420}) have some common data. In particular, the \texttt{side} variable tells in which detector (left:-1 or right:+1 of the interaction point) the hit has been seen. Moreover, some generator level data is provided for information, as the correspondance with the contents of the \texttt{GEN} tree is not possible. These generator-level data correspond to the particle kinematics (energy, momentum, angle) and identification (pid).

\begin{quote}
\begin{tabular}{ll}
\multicolumn{2}{l}{\textbf{Common leaves for ZDC, RP220, FP420}}\\
   \texttt{~~~float T } &\texttt{ // time of flight  in s }\\
   \texttt{~~~float E } &\texttt{ // measured/smeared energy in GeV }\\
   \texttt{~~~int side }&\texttt{ // -1 or +1 }\\
\multicolumn{2}{l}{Generator level data}\\
   \texttt{~~~int pid;     }&\texttt{ // particle ID }\\
   \texttt{~~~float genPx;    }&\texttt{ // particle momentum vector (x component) in GeV$/c$ }\\
   \texttt{~~~float genPy;    }&\texttt{ // particle momentum vector (y component) in GeV$/c$ }\\
   \texttt{~~~float genPz;    }&\texttt{ // particle momentum vector (z component) in GeV$/c$ }\\
   \texttt{~~~float genPT;    }&\texttt{ // particle transverse momentum in GeV$/c$ }\\
   \texttt{~~~float genEta;   }&\texttt{ // particle pseudorapidity   }\\
   \texttt{~~~float genPhi;   }&\texttt{ // particle azimuthal angle in rad }\\
\end{tabular}
\end{quote}

\begin{quote}
\begin{tabular}{ll}
\multicolumn{2}{l}{\textbf{Additional leaves in the \texttt{ZDChits} branch (\texttt{Analysis} tree)}}\\
   \texttt{~~~int hadronic\_hit} &\texttt{// 0(is not hadronic) or 1(is hadronic) }
\end{tabular}
\end{quote}

\begin{quote}
\begin{tabular}{ll}
\multicolumn{2}{l}{\textbf{Additional leaves in the \texttt{RP220hits} and \texttt{FP420hits} branches (\texttt{Analysis} tree)}}\\
   \texttt{~~~flaot S } &\texttt{ // detector position from IP in m } \\
   \texttt{~~~float X } &\texttt{ // hit horizontal position in m } \\
   \texttt{~~~float Y } &\texttt{ // hit vertical position in m }   \\
   \texttt{~~~float TX } &\texttt{ // hit horizontal angle in rad } \\
   \texttt{~~~float TY } &\texttt{ // hit vertical angle in rad } \\
   \texttt{~~~float q2 } &\texttt{ // reconstructed momentum transfer in GeV$^2$ }
\end{tabular}
\end{quote}
The hit position is computed from the center of the beam position, not from the edge of the detector.

\subsection{Deeper description of jet algorithms}

In this section, we briefly describe the differences between the six jet algorithms interfaced in \textit{Delphes}, via the FastJet utiliy~\citep{bib:FASTJET}. Jet algorithms differ in their sensitivity to soft particles or collinear splittings, and in their computing speed performances. The first three belong to the cone algorithm class while the last three are using a sequential recombination scheme. For all of them, the calorimetric cells are used as inputs for the jet clustering. 
 
\subsubsection*{Cone algorithms}
 
\begin{enumerate}
 
\item {\it CDF Jet Clusters}~\citep{bib:jetclu}: Basic cone reconstruction algorithm used by the \textsc{CDF} experiment in Run II). All cells lying in a circular cone around the jet axis with a transverse energy $E_T$ higher than a given threshold are used to seed the jet candidates. This algorithm is fast but sensitive to both soft particles and collinear splittings. 

\item {\it CDF MidPoint}~\citep{bib:midpoint}: Cone reconstruction algorithm developed for the \textsc{CDF} Run II to reduce infrared and collinear sensitivities compared to purely seed-based cone by adding `midpoints' (energy barycentres) in the list of cone seeds.
 
\item {\it Seedless Infrared Safe Cone}~\citep{bib:SIScone}: The \textsc{SISC}one algorithm is simultaneously insensitive to additional soft particles and collinear splittings, and fast enough to be used in experimental analysis.
 
\end{enumerate}

\subsubsection*{Recombination algorithms}
 
The three sequential recombination jet algorithms are safe with respect to soft radiations (\textit{infrared}) and collinear splittings. They rely on recombination schemes where calorimeter cell pairs are successively merged. 
The definitions of the jet algorithms are similar except for the definition of the \textit{distances} $d$ used during the merging procedure. Two such variables are defined: the distance $d_{ij}$ between each pair of cells $(i,j)$, and a variable $d_{iB}$ (\textit{beam distance}) depending on the transverse momentum of the cell $i$.
The jet reconstruction algorithm browses the calorimetric cell list. It starts by finding the minimum value $d_\textrm{min}$ of all the distances $d_{ij}$ and $d_{iB}$. If $d_\textrm{min}$ is a $d_{ij}$, the cells $i$ and $j$ are merged into a single cell with a four-momentum $p^\mu = p^\mu (i) + p^\mu (j)$ (\textit{E-scheme recombination}). If $d_\textrm{min}$ is a $d_{iB}$, the cell is declared as a final jet and is removed from the input list. This procedure is repeated until no cells are left in the input list. Further information on these jet algorithms is given here below, using $k_{ti}$, $y_{i}$ and $\phi_i$ as the transverse momentum, rapidity and azimuth of calorimetric cell $i$ and $\Delta R_{ij}= \sqrt{(y_i-y_j)^2+(\phi_i-\phi_j)^2}$ as the jet-radius parameter:
 
\begin{enumerate}[start=4]
 
\item {\it Longitudinally invariant $k_t$ jet}~\citep{bib:ktjet}, with 
  $d_{ij} = \min(k_{ti}^2,k_{tj}^2) \times \frac{\Delta R_{ij}^2}{R^2}$ and $d_{iB}=k_{ti}^2$,
\item {\it Cambridge/Aachen jet}~\citep{bib:aachen}, with $d_{ij} = \frac{\Delta R_{ij}^2}{R^2}$ and $d_{iB}=1$,
\item {\it Anti $k_t$ jet}~\citep{bib:antikt}, where hard jets are exactly circular in the $(y,\phi)$ plane:
$d_{ij} =  \min(1/k_{ti}^2,1/k_{tj}^2) \times \frac{\Delta R_{ij}^2}{R^2}$ and $d_{iB}=\frac{1}{k_{ti}^2}$.
\end{enumerate}

\subsection{Running an analysis on your \textit{Delphes} events}
 
To analyse the \textsc{ROOT} ntuple produced by \textit{Delphes}, the simplest way is to use the {\verb Analysis_Ex.cpp } code which is coming in the {\verb Examples } repository of \textit{Delphes}. Note that all of this is optional and done to facilitate the analyses, as the output from \textit{Delphes} is viewable with the standard \textsc{ROOT} \texttt{TBrowser} and can be analysed using the \texttt{MakeClass} facility.
As an example, here is a simple overview of a \texttt{myoutput.root} file created by \textit{Delphes}:
\begin{quote}
\begin{verbatim}
me@mylaptop:~$ root -l myoutput.root
root [0]
Attaching file myoutput.root as _file0...
root [1] .ls
TFile**         myoutput.root
 TFile*         myoutput.root
  KEY: TTree    GEN;1   Analysis tree
  KEY: TTree    Analysis;1      Analysis tree
  KEY: TTree    Trigger;1       Analysis tree
root [2] TBrowser t;
root [3] Analysis->GetEntries()
(const Long64_t)200
root [4] GEN->GetListOfBranches()->ls()
OBJ: TBranchElement  Event       Event_ : 0 at: 0x9108f30
OBJ: TBranch         Event_size  Event_size/I : 0 at: 0x910cfd0
OBJ: TBranchElement  Particle    Particle_ : 0 at: 0x910c6b0
OBJ: TBranch  Particle_size  Particle_size/I : 0 at: 0x9111c58
root [5] Trigger->GetListOfLeaves()->ls()
OBJ: TLeafElement TrigResult_          TrigResult_ : 0 at: 0x90f90a0
OBJ: TLeafElement TrigResult.Accepted  Accepted[TrigResult_] : 0 at: 0x90f9000
OBJ: TLeafI       TrigResult_size      TrigResult_size : 0 at: 0x90fb860
\end{verbatim}
\end{quote}
The \texttt{.ls} command lists the current keys available and in particular the three \textit{tree} names.
\mbox{\texttt{TBrowser t}} launches a browser and the \texttt{GetEntries()} method outputs the number of data in the corresponding \textit{tree}.
The list of \textit{branches} or \textit{leaves} can be displayed with the \texttt{GetListOfBranches()} and \texttt{GetListOfLeaves()} methods, pointing to the \texttt{ls()} one. In particular, it is possible to shown only parts of the output, using wildcard characters (\texttt{*}):
\begin{quote}
\begin{verbatim}
root [6] Analysis->GetListOfLeaves()->ls("*.E")
OBJ: TLeafElement       Jet.E           E[Jet_] : 0 at: 0xa08bc68
OBJ: TLeafElement       TauJet.E        E[TauJet_] : 0 at: 0xa148910
OBJ: TLeafElement       Electron.E      E[Electron_] : 0 at: 0xa1d8a50
OBJ: TLeafElement       Muon.E          E[Muon_] : 0 at: 0xa28ac80
OBJ: TLeafElement       Photon.E        E[Photon_] : 0 at: 0xa33cd88
OBJ: TLeafElement       Tracks.E        E[Tracks_] : 0 at: 0xa3cced0
OBJ: TLeafElement       CaloTower.E     E[CaloTower_] : 0 at: 0xa4ba188
OBJ: TLeafElement       ZDChits.E       E[ZDChits_] : 0 at: 0xa54a3c8
OBJ: TLeafElement       RP220hits.E     E[RP220hits_] : 0 at: 0xa61e648
OBJ: TLeafElement       FP420hits.E     E[FP420hits_] : 0 at: 0xa6d0920
\end{verbatim}
\end{quote}

To draw a particular leaf, either double-click on the corresponding name in the \texttt{TBrowser} or use the \texttt{Draw} method of the corresponding \textit{tree}.
\begin{quote}
\begin{verbatim}
root [7] Trigger->Draw("TrigResult.Accepted");
\end{verbatim}
\end{quote}
Mathematical operations on several \textit{leaves} are possible within a given \textit{tree}, following the C++ syntax:
\begin{quote}
\begin{verbatim}
root [8] Analysis->Draw("Muon.Px * Muon.Px");
root [9] Analysis->Draw("sqrt(pow(Muon.E,2) -  pow(Muon.Pz,2) + pow(Muon.PT,2))");
\end{verbatim}
\end{quote}
Finally, to prepare an deeper analysis, the \texttt{MakeClass} method is useful. It creates two files (\texttt{*.h} and \texttt{*.C}) with automatically generated code that allows the access to all branches and leaves of the corresponding tree:
\begin{quote}
\begin{verbatim}
root [10] Trigger->MakeClass()
Info in <TTreePlayer::MakeClass>: Files: Trigger.h and 
        Trigger.C generated from TTree: Trigger
\end{verbatim}
\end{quote}
For more information, refer to ROOT documentation. Moreover, an example of code (based on the output of \texttt{MakeClass}) is provided in the \texttt{Examples/} directory.

To run the \texttt{Examples/Analysis\_Ex.cpp} code, the two following arguments are required: a text file containing the input \textit{Delphes} \texttt{root} files to run, and the name of the output \texttt{root} file.
 \begin{quote}
\begin{verbatim}
me@mylaptop:~$ ./Analysis_Ex input_file.list output_file.root
\end{verbatim}
 \end{quote}
One can easily edit, modify and compile (\texttt{make}) changes in this file.

\subsubsection{Adding the trigger information}
The \texttt{Examples/Trigger\_Only.cpp} code permits to run the trigger selection separately from the general detector simulation on output \textit{Delphes} root files. A \textit{Delphes} \texttt{root} file is mandatory as an input argument for the \texttt{Trigger\_Only} routine. The new \textit{tree} containing the trigger result data will be appended to this file.
The trigger datacard is also necessary. To run the code:
 \begin{quote}
\begin{verbatim}
me@mylaptop:~$ ./Trigger_Only input_file.root data/TriggerCard.dat
\end{verbatim}
 \end{quote}
 
\subsection{Running the FROG event display}
 
\begin{itemize}
\item If the { \verb FLAG_FROG } was switched on in the smearing card, two files have been created during the running of \textit{Delphes}: \texttt{DelphesToFROG.vis} and \texttt{DelphesToFROG.geom }. They contain all the needed pieces of information to run \textsc{FROG}.
\item To display the events and the geometry, you first need to compile \textsc{FROG}. Go to the {\verb Utilities/FROG } and type {\verb make }. This compilation is done once for all, with this geometry (i.e.\ as long as the \texttt{*vis} and \texttt{*geom} files do not change).
\item Go back into the main directory and type 
\begin{quote}
\texttt{me@mylaptop:~\$ ./Utilities/FROG/FROG}
\end{quote}
\end{itemize}

\subsection{LHCO file format}
 The \texttt{*LHCO} file format is a text-\textsc{ASCII} data format briefly discussed here. An exhaustive description is provided on \href{http://v1.jthaler.net/olympicswiki}{http://v1.jthaler.net/olympicswiki}. This section is based on this webpage.
Only final high-level objects are available in the \texttt{LHCO} format, and their properties are arranged in columns. Each row corresponds to an object in the event and all events are written after each other. Comment-lines starts with a hash \texttt{\#} symbol.

\begin{verbatim}
  #  typ   eta     phi     pt     jmas    ntrk    btag   had/em    dum1   dum2
  0          57     0
  1   0   1.392  -2.269  19.981   0.000   0.000   0.000   4.605   0.000  0.000
  2   3   1.052   2.599  29.796   3.698  -1.000   0.000   0.320   0.000  0.000
  3   4   1.542  -2.070  84.308  41.761   7.000   0.000   1.000   0.000  0.000
  4   4   1.039   0.856  58.992  34.941   1.000   0.000   1.118   0.000  0.000
  5   4   1.052   2.599  29.796   3.698   0.000   0.000   0.320   0.000  0.000
  6   4   0.431  -2.190  22.631   3.861   0.000   0.000   1.000   0.000  0.000
  7   6   0.000   0.845  62.574   0.000   0.000   0.000   0.000   0.000  0.000
\end{verbatim}
Each row in an event starts with a unique number (i.e.\ in first column). 
Row \texttt{0} contains the event number (here: \texttt{57}) and some trigger information (here: \texttt{0}. This very particular trigger encoding is not implemented in \textit{Delphes}.).
Subsequent rows list the reconstructed high-level objects. 
Each row is organised in columns, which details the object kinematics as well as more specific information, such as isolation criteria or $b$-tagging.

\paragraph{1st column (\texttt{\#})}
The first column is the line number in the event. Each event starts with a 0 and contains as many lines as needed to list all high-level objects.

\paragraph{2nd column (\texttt{typ})}
The second column gives the object identification code, or \textit{type}. 
The different object types are:\\
\begin{tabular}{ll}
  \texttt{0}& for a photon ($\gamma$)\\
  \texttt{1}& for an electron ($e^\pm$)\\
  \texttt{2}& for a muon ($\mu^\pm$)\\
  \texttt{3}& for a hadronically-decaying tau ($\tau$-jet)\\
  \texttt{4}& for a jet\\
  \texttt{6}& for a missing transverse energy ($E_T^\textrm{miss}$)\\
\end{tabular}\\
Object type \texttt{5} is not defined.
An event always ends with the row corresponding to the missing transverse energy (type \texttt{6}).

\paragraph{3rd (\texttt{eta}) and 4th (\texttt{phi}) columns}
The third and forth columns gives the object pseudorapidity $\eta$ and azimuth $\phi$. This latter quantity is expressed in radians, ranging from $-\pi$ to $\pi$.

\paragraph{5th (\texttt{pt}) and 6th (\texttt{jmass}) columns}
The fifth column provides the object transverse momentum ($p_T$ in GeV$/c$) or energy ($E_T$ in GeV), while the invariant mass ($M$ in GeV/$c^2$) is in the sixth column.

\paragraph{7th column (\texttt{ntrk})}
The seventh column reports the total number of tracks associated to the objects. This is \texttt{0} for photons, \texttt{$\pm$ 1} for charged leptons including taus (where the sign reports the lepton measured charge) and a positive number (\texttt{$\geq$ 0}) for jets.

\paragraph{8th column (\texttt{btag})}   
The eighth column tells whether a jet is tagged as a $b$-jet (\texttt{1}) or not (\texttt{0}).
This is always \texttt{0} for electrons, photons and missing transverse energy.
For muons, the closest jet in searched for, in terms of $\Delta R$. The integer-part of the quoted number is the row-number (column 1) of this jet.

\paragraph{9th column (\texttt{had/em})}
For jets, electrons and photons, the ninth column is the ration between hadronic and electromagnetic energies in the calorimetric cells associated to the object. This is always \texttt{0} for missing transverse energy.
For muons, this number (\texttt{aaa.bb}) reports two values related to the muon isolation (section \ref{sec:isolation}). The integer part (\texttt{aaa}) is transverse momentum sum $P_T$ (in GeV/$c$) and the fractional part (\texttt{bb}) is the energy ratio $\rho_\mu$.

\paragraph{10th and 11th columns (\texttt{dum1} and \texttt{dum2})}
The last two columns are currently not used.

\paragraph{Warning}
Inherently to the data format itself, the \texttt{*LHCO} output contains only a fraction of the available data. Moreover, dealing with text file may have various drawbacks, such as the output file size and the time needed for its creation. Whenever possible, working on the \texttt{*root} output file should be preferred.


\begin{thebibliography}{99}
\addcontentsline{toc}{section}{References}

\bibitem{bib:geant} J. Allison, et al., Nucl. Inst. \& Meth. in \textbf{Phys. Res. A} \href{http://dx.doi.org/10.1016/S0168-9002(03)01368-8}{506 (2003) 250-303}; \textbf{IEEE Trans. on Nucl. Sc.} \href{http://dx.doi.org/10.1109/TNS.2006.869826}{53:1 (2006) 270-278}.
  
\bibitem{bib:delphes} \textit{Delphes}, \href{http://www.fynu.ucl.ac.be/delphes.html}{www.fynu.ucl.ac.be/delphes.html}
\bibitem{bib:Root} 
R. Brun, F. Rademakers, Nucl. Inst. \& Meth. in \textbf{Phys. Res. A} \href{http://dx.doi.org/10.1016/S0168-9002(97)00048-X}{389 (1997) 81-86}.
\bibitem{bib:ExRootAnalysis} 
P. Demin, (2006), unpublished. Now part of MadGraph/MadEvent.
\bibitem{bib:cmsjetresolution} The \textsc{CMS} Collaboration, \textbf{CERN/LHCC} \href{http://documents.cern.ch/cgi-bin/setlink?base=LHCc&categ=public&id=LHCc-2006-001}{2006-001}.
\bibitem{bib:ATLASresolution} The \textsc{ATLAS} Collaboration, \textbf{CERN-OPEN} 2008-020, \\arXiv:\href{http://arxiv.org/abs/arxiv:0901.0512}{0901.0512v1}[hep-ex].
\bibitem{bib:hector} 
X. Rouby, J. de Favereau, K. Piotrzkowski, \textbf{JINST} \href{http://www.iop.org/EJ/abstract/1748-0221/2/09/P09005}{2 P09005 (2007)}.
\bibitem{bib:FASTJET} 
M. Cacciari, G.P. Salam, \textbf{Phys. Lett. B} \href{http://dx.doi.org/10.1016/j.physletb.2006.08.037}{641 (2006) 57}.
\bibitem{bib:jetclu} 
F. Abe et al. (CDF Coll.), \textbf{Phys. Rev. D} \href{http://link.aps.org/doi/10.1103/PhysRevD.45.1448}{45 (1992) 1448}.
\bibitem{bib:midpoint} 
G.C. Blazey, et al., arXiv:\href{http://arxiv.org/abs/hep-ex/0005012}{0005012}[hep-ex].
\bibitem{bib:SIScone} 
G.P. Salam, G. Soyez, \textbf{JHEP} \href{http://dx.doi.org/10.1088/1126-6708/2007/05/086}{05 (2007) 086}.
\bibitem{bib:ktjet} S. Catani, Y.L. Dokshitzer, M.H. Seymour, B.R. Webber, \textbf{Nucl. Phys. B} \href{http://dx.doi.org/10.1016/0550-3213(93)90166-M}{406 (1993) 187}; S.D. Ellis, D.E. Soper, \textbf{Phys. Rev. D} \href{http://link.aps.org/doi/10.1103/PhysRevD.48.3160}{48 (1993) 3160}.
\bibitem{bib:aachen} Y.L. Dokshitzer, G.D. Leder, S. Moretti, B.R. Webber, \textbf{JHEP} \href{http://dx.doi.org/10.1088/1126-6708/1998/01/011}{08} \href{http://dx.doi.org/10.1088/1126-6708/1998/01/011}{(1997) 001}; M. Wobisch, T. Wengler, arXiv:\href{http://arxiv.org/abs/hep-ph/9907280}{9907280}[hep-ph].
\bibitem{bib:antikt} 
M. Cacciari, G.P. Salam, G. Soyez, \textbf{JHEP} \href{http://dx.doi.org/10.1088/1126-6708/2008/04/063}{04 (2008) 063}.
\bibitem{bib:pdg} C. Amsler et al. (Particle Data Group), \textbf{Phys. Lett. B} \href{http://dx.doi.org/10.1016/j.physletb.2008.07.018}{667 (2008) 1}.
\bibitem{bib:whphotoproduction} S. Ovyn, \textbf{Nucl. Phys. Proc. Suppl.} \href{http://dx.doi.org/10.1016/j.nuclphysbps.2008.07.034}{179-180 (2008) 269-276}.
\bibitem{bib:mgme} 
J. Alwall, et al., \textbf{JHEP} \href{http://dx.doi.org/10.1088/1126-6708/2007/09/028}{09 (2007) 028}.
\bibitem{bib:pythia} 
T. Sjostrand, S. Mrenna, P. Skands, \textbf{JHEP} \href{http://dx.doi.org/10.1088/1126-6708/2006/05/026}{05 (2006) 026}.
\bibitem{bib:cmstauresolution} 
R. Kinnunen, A.N. Nikitenko, \textbf{CMS NOTE} \href{http://cdsweb.cern.ch/record/687274}{1997/002}.
\bibitem{bib:FROG} L. Quertenmont, V. Roberfroid, \textbf{CMS CR} \href{http://cms.cern.ch/iCMS/jsp/openfile.jsp?type=CR&year=2009&files=CR2009_028.pdf}{2009/028}, arXiv:\href{http://arxiv.org/abs/0901.2718}{0901.2718v1}[hep-ex].
\bibitem{bib:wtphotoproduction} J. de Favereau de Jeneret, S. Ovyn, 
\textbf{Nucl. Phys. Proc. Suppl.}
\href{http://dx.doi.org/10.1016/j.nuclphysbps.2008.07.040}{179-180 (2008)}
\href{http://dx.doi.org/10.1016/j.nuclphysbps.2008.07.040}{277-284}; S. Ovyn, J.
de Favereau de Jeneret, \href{http://dx.doi.org/10.1393/ncb/i2008-10684-5}{Nuovo
Cimento B}, arXiv:0806.4841[hep-ph].

\bibitem{bib:papierquisortirajamais}J. de Favereau~et~al, \href{http://arxiv.org/abs/0908.2020}{arXiv:0908.2020v1} [hep-ph] (2008), to be published in EPJ.

\bibitem{bib:papiersimon} S. de Visscher,  J.M. Gerard, M. Herquet, V.
Lema\^itre, F. Maltoni, \textbf{JHEP}
\href{http://dx.doi.org/10.1088/1126-6708/2009/08/042}{08 (2009) 042}.

\bibitem{bib:mcfio} P. Lebrun, L. Garren, Copyright (c) 1994-1995 Universities Research Association, Inc.
\bibitem{bib:stdhep} L.A. Garren, M. Fischler, \href{http://cepa.fnal.gov/psm/stdhep/c++}{cepa.fnal.gov/psm/stdhep/c++}
\bibitem{bib:hepmc} M. Dobbs and J.B. Hansen, \textbf{Comput. Phys. Commun.} \href{http://dx.doi.org/10.1016/S0010-4655(00)00189-2}{134 (2001) 41}.
\bibitem{bib:lhe} J. Alwall, et al., \textbf{Comput. Phys. Commun.} \href{http://dx.doi.org/10.1016/j.cpc.2006.11.010}{176:300-304,2007}.

\end{thebibliography}

\begin{thebibliography}{2}
\addcontentsline{toc}{section}{Internal code references}

\bibitem[a]{qr:inputformat} The standard Monte Carlo event structures \texttt{StdHEP}~\citep{bib:stdhep} and \texttt{HepMC}~\citep{bib:hepmc} can be used as an input. Besides, \textit{Delphes} can also provide detector response for events read in ``Les Houches Event Format'' (LHEF~\citep{bib:lhe}) and \texttt{*.root} files obtained from \texttt{*.hbook} using the \texttt{h2root} utility from the ROOT framework~\citep{bib:Root}.
See the following classes: \texttt{HEPEVTConverter}, \texttt{HepMCConverter}, \texttt{LHEFConverter}, \texttt{STDHEPConverter} and \texttt{DelphesRootConverter}. 

\bibitem[b]{qr:outputformat} The ROOT output files are created using the \texttt{ExRootAnalysis} utility~\citep{bib:ExRootAnalysis}. Generator-level data are located under the \texttt{GEN} tree, the analysis data objects after reconstruction under the \texttt{Analysis} tree, and the results of the trigger emulation under the \texttt{Trigger} tree. 

\bibitem[c]{qr:lhco} Set the \texttt{FLAG\_LHCO} variable to $1$ or $0$ in the detector card to switch on/off the creation of \texttt{*.lhco} output file.

\bibitem[d]{qr:invisibleparticles} The list of particles considered as invisible is accessible in the \texttt{PdgParticle} class. This list currently contains the PIDs 12, 14, 16, 1000022,  1000023, 1000025, 1000035 and 1000045, in absolute values.

\bibitem[e]{qr:detectorcard}The detector card is the \texttt{data/DetectorCard.dat} file. This file is parsed by the \texttt{SmearUtil} class.

\bibitem[f]{qr:datacards} Detector and trigger cards for the \textsc{ATLAS} and \textsc{CMS} experiments are also provided in \texttt{data/} directory.

\bibitem[g]{qr:resolutionterms}The resolution terms in the detector card are named \texttt{ELG\_Xyyy} or \texttt{HAD\_Xyyy},  refering to electromagnetic and hadronic terms (resp.); \texttt{X} is replaced by \texttt{S}, \texttt{N}, \texttt{C} for the stochastic, noise and constant terms; and finally \texttt{yyy} is \texttt{cen} for central part, \texttt{ec} for end-caps, \texttt{fwd} for the forward calorimeters and \texttt{zdc} for the zero-degree calorimeters. 

\bibitem[h]{qr:magneticfield} See the \texttt{TrackPropagation} class.

\bibitem[i]{qr:tracks} See the \texttt{TRACK\_eff} and \texttt{TRACK\_ptmin} terms in the detector card.

\bibitem[j]{qr:energysmearing} The response of the detector is applied to the electromagnetic and the hadronic particles through the \texttt{SmearElectron} and \texttt{SmearHadron} methods in the \texttt{SmearUtil} class.

\bibitem[k]{qr:emhadratios} To implement different ratios for other particles, see the \texttt{BlockClasses} class.

\bibitem[l]{qr:calorimetriccells} As the detector is assumed to be cylindrical (e.g.\ symmetric in $\phi$ and with respect to the $\eta=0$ plane), the detector card stores the number of calorimetric cells with $\phi=0$ and $\eta>0$ (default: $40$ cells). For a given $\eta$, the size of the $\phi$ segmentation is also specified. See the \texttt{TOWER\_number}, \texttt{TOWER\_eta\_edges} and \texttt{TOWER\_dphi} variables in the detector card.

\bibitem[m]{qr:analysistree} All these processed data are located under the \texttt{Analysis} tree.

\bibitem[n]{qr:muonsmearing} See the \texttt{SmearMuon} method in the \texttt{SmearUtil} class.

\bibitem[o]{qr:isolflag} See the \texttt{IsolFlag} and \texttt{IsolPt} values in the \texttt{Electron} or \texttt{Muon} collections in the \texttt{Analysis} tree, as well as the \texttt{ISOL\_PT} and \texttt{ISOL\_Cone} variables in the detector card.

\bibitem[p]{qr:caloisolation} Calorimetric isolation parameters in the detector card are \texttt{ISOL\_Calo\_ET} and  \texttt{ISOL\_Calo\_Grid} in the detector card.

\bibitem[q]{qr:fwdneutrals} These thresholds are defined by the \texttt{ZDC\_gamma\_E} and \texttt{ZDC\_n\_E} variables in the detector card.

\bibitem[r]{qr:jetalgo} The choice is done by allocating the \texttt{JET\_jetalgo } input parameter in the detector card.

\bibitem[s]{qr:ptcutjet} See the \texttt{PTCUT\_jet }variable in the detector card. 

\bibitem[t]{qr:jetparams} See the \texttt{JET\_coneradius} and \texttt{JET\_seed} variables in the detector card. The existing FastJet code has been modified to allow easy modification of the cell pattern in $(\eta, \phi)$ space.
In following versions of \textit{Delphes}, a new dedicated plug-in will be created on this purpose.

\bibitem[u]{qr:energyflow} Set \texttt{JET\_Eflow} to $1$ or $0$ in the detector card in order to switch on or off the energy flow for jet reconstruction.

\bibitem[v]{qr:btag} Corresponding to the \texttt{BTAG\_b}, \texttt{BTAG\_mistag\_c} and \texttt{BTAG\_mistag\_l} constants, for the efficiency of tagging of a $b$-jet, the efficiency of mistagging a $c$-jet as a $b$-jet, and the 
efficiency of mistagging a light jet ($u$,$d$,$s$,$g$) as a $b$-jet.

\bibitem[w]{qr:taujets} See the following parameters in the detector card:\\
\texttt{TAU\_energy\_scone } for $R^\textrm{em}$; \texttt{JET\_M\_seed }  for min $E_{T}^\textrm{cell}$;
\texttt{TAU\_energy\_frac} for $C_{\tau}$; \texttt{TAU\_track\_scone} for $R^\textrm{tracks}$; 
 \texttt{PTAU\_track\_pt } for min $p_T^\textrm{tracks}$ and \texttt{TAUJET\_pt} for $\min p_T$.


\bibitem[x]{qr:triggercard} The trigger card is the \texttt{data/TriggerCard.dat} file. Default trigger files are also available for CMS-like and ATLAS-like detectors

\bibitem[y]{qr:protontaggers} The resolution is defined by the \texttt{RP220\_T\_resolution} and \texttt{RP420\_T\_resolution} parameters in the detector card.

\bibitem[z]{qr:frog} To prepare the visualisation, the \texttt{FLAG\_FROG} parameter should be equal to $1$.

\end{thebibliography}
\end{document}